\documentclass[nofootinbib,twocolumn,showpacs,preprintnumbers,pre,aps]{revtex4-1}

\usepackage{graphicx}
\usepackage{bm}
\usepackage{amsmath}
\usepackage{amssymb}
\usepackage{color}

\begin{document}

\title{Anomalous diffusion in viscoelastic media with active force dipoles}%

\author{Kento Yasuda}

\author{Ryuichi Okamoto}

\author{Shigeyuki Komura}\email{komura@tmu.ac.jp}

\affiliation{
Department of Chemistry, Graduate School of Science and Engineering,
Tokyo Metropolitan University, Tokyo 192-0397, Japan}

\date{\today}

\begin{abstract}
With the use of the ``two-fluid model", we discuss anomalous diffusion induced by active force 
dipoles in viscoelastic media. 
Active force dipoles, such as proteins and bacteria, generate non-thermal fluctuating flows that 
lead to a substantial increment of the diffusion.
Using the partial Green's function of the two-fluid model, we first obtain passive (thermal) two-point 
correlation functions such as the displacement cross-correlation function between the two 
point particles separated by a finite distance. 
We then calculate active (non-thermal) one-point and two-point correlation functions due to 
active force dipoles.
The time correlation of a force dipole is assumed to decay exponentially with a characteristic 
time scale.  
We show that the active component of the displacement cross-correlation function exhibits various 
crossovers from super-diffusive to sub-diffusive behaviors depending on the  characteristic time 
scales and the particle separation.
Our theoretical results are intimately related to the microrheology technique to detect fluctuations 
in non-equilibrium environment.
\end{abstract}

\maketitle

\section{Introduction}
\label{sec:introduction}

The cytoplasm of living cells is full of proteins and organelles that play important active 
roles with the aid of chemical fuels such as adenosine triphosphate 
(ATP)~\cite{AlbertsBook}. 
In such a non-equilibrium environment, the transport properties of chemical species 
drastically deviate from those in static equilibrium conditions.  
For example, there are several experimental works reporting the anomalous diffusion 
of a tagged particle in biological cells due to protein 
activities~\cite{Brangwynne08,MacKintosh12,Weber12,Parry14,Guo14}.
In other systems, a large enhancement of diffusion was also observed for a passive particle 
immersed in a bacterial bath~\cite{Wu00,Chen07} or in a suspension of algae 
Chlamydomonas~\cite{Leptos09}, and such a phenomenon has been also studied
theoretically~\cite{Hatwalne04,Underhill11}.

The modified diffusion in cells was attributed to non-equilibrium forces generated by molecular 
motors walking on cytoskeletal networks~\cite{Mac08,Levine09}. 
Recently, Mikhailov and Kapral proposed a different mechanism caused by non-equilibrium 
conformational changes of proteins or enzymes~\cite{Mikhailov15,Kapral16}. 
They showed that, in addition to thermal fluctuations, active proteins in living cells generate
non-thermal fluctuating flows that lead to a substantial increment of the diffusion constant.
A chemotaxis-like drift of a passive particle was also predicted when a spatial gradient of active 
proteins is present~\cite{Mikhailov15,Kapral16}. 
In these previous works, however, the three-dimensional (3D) cytoplasm and two-dimensional 
(2D) biomembrane were treated as purely viscous fluids characterized by constant shear 
viscosities~\cite{Koyano16}.

In general, biological cells behave as viscoelastic materials~\cite{Fabry01,Kollmannsberger11}.
Hoffman \textit{et al}.\ experimentally determined the frequency-dependent shear modulus 
of cultured mammalian cells by using various methods to measure their  
viscoelastic properties~\cite{Hoffman06,Hoffman09}.
Interestingly, they found two universal (weak) power-law dependencies of the shear modulus
at low frequencies corresponding to the cortical and intracellular networks.
At high frequencies, on the other hand, they observed an exponent of 3/4 which was attributed 
to the mechanical response of actin fibers.
Such an universal behavior of mechanical responses in living cells was also reported in other
work~\cite{Trepat07}.

Among various methods, microrheology is one of the most useful techniques to 
measure the rheological properties of living cells~\cite{MW95,MGZWK,GSOMS,Schnurr97,Mason00}.
In this method, the local and bulk mechanical properties of a single cell can be extracted from 
a Brownian motion of probe particles, including both thermal and non-thermal contributions~\cite{SM10}.
Concerning its theoretical background, the generalized Stokes-Einstein relation (GSER), 
equivalent to the fluctuation dissipation theorem (FDT), has been used to analyze thermal diffusive motions.
In non-thermal situations, the GSER has been further extended to relate particle mean squared 
displacement (MSD) and non-thermal force fluctuations~\cite{Lau03,Lau09}. 
It should be noted, however, that the GSER contains various assumptions which can be violated in 
several situations~\cite{SM10}. 
Therefore it is necessary to discuss both thermal and non-thermal Brownian motions in a 
viscoelastic medium which is described by a well-founded theoretical model.

In this paper, we discuss diffusive motion of passive particles embedded in viscoelastic media 
that is described by the ``two-fluid model" for gels~\cite{deGennes76a,deGennes76b,Brochard77}. 
We especially focus on the effects of non-thermal fluctuations induced by active force dipoles
which undergo cyclic motions. 
We calculate displacement cross-correlation functions (CCF) of two point particles for the passive
situation induced by thermal fluctuations and the active situation driven by force dipole fluctuations.
Our calculation is closely related to the ``two-point microrheology" method which has several 
technical advantages compared to the ``one-point microrheology"~\cite{CVWGKYW}.
As for the stochastic property of a force dipole, we consider the case when there is no 
correlation between different times, and also the case when it decays exponentially with a 
characteristic time scale. 
If the dipole time scale is much larger than the viscoelastic time scale, we show that the active 
contribution of the displacement CCF exhibits all the possible crossover behaviors between 
super-diffusive and sub-diffusive motions.  
Our predictions can be applied not only for cells but also for bacterial suspensions and systems 
containing active colloids.

Since our theory is based on the standard two-fluid model, it has some similarities to the works 
by Levine and Lubensky~\cite{Levine00,Levine01} or MacKintosh and 
Levine~\cite{Mac08,Levine09}. 
In the former studies~\cite{Levine00,Levine01}, they investigated the dynamics of rigid spheres 
embedded in viscoelastic media by using the two-fluid model, but did not consider the effects of 
non-thermal fluctuations. 
In the latter studies~\cite{Mac08,Levine09}, on the other hand, they developed a model for 
F-actin networks driven out of equilibrium by molecular motors.
The main difference in our work is that active force dipoles are embedded in the fluid and exert 
forces on the fluid itself. 
In this regard, we use the partial Green's function that connects the force acting on the fluid and the 
fluid velocity as discussed in Refs.~\cite{Sonn14,Sonn14-2}. 
In these works, they emphasized the role of the intermediate length scale in the analysis 
of microrheology data. 
In our separate work, starting from the two-fluid model, we have derived effective equations of 
motions for tracer particles displaying local deformations and local fluid flows~\cite{Yasuda17}.

In the next section, we describe the two-fluid model and show its partial Green's function both 
in the Fourier space and the real space. 
In Sec.~\ref{sec:pass2pcorr}, we discuss the passive two-point correlation functions.
Using the coupling mobilities and the FDT in thermal equilibrium, we calculate the power spectral 
density of the velocity CCFs and the displacement CCFs.
In Sec.~\ref{sec:1pcorr}, we shall investigate active one-point correlation functions due to active force 
dipoles.
We calculate the active velocity auto-correlation function of a passive point particle by assuming 
different time correlations of force dipoles. 
We then discuss in Sec.~\ref{sec:act2pcorr} the active two-point correlation functions which are
useful for two-point microrheology.
The summary of our work and some discussions related to the recent experiments are given 
in Sec.~\ref{sec:discussion}.

\begin{figure}[tbh]
\begin{center}
\includegraphics[scale=0.35]{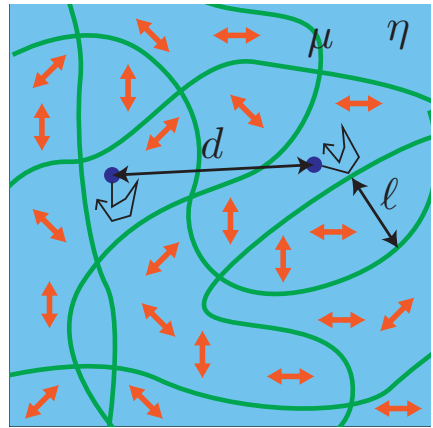}
\end{center}
\caption{
Schematic representation of the two-fluid model. 
The system consists of an elastic network characterized by the Lam\'e coefficient $\mu$,
and a viscous fluid characterized by the shear viscosity $\eta$. 
The elastic and fluid components are coupled through the mutual friction. 
The length scale $\ell$ characterizes the typical internal structure of the elastic network. 
Orange objects represent stochastic force dipoles which are immersed in the fluid component.
Two passive point particles separated by a distance $d$ are embedded in the fluid component. 
These passive particles undergo correlated random Brownian motion due to thermal fluctuations
and active stochastic fluctuations induced by active force dipoles.
}
\label{two-fluid}
\end{figure}

\section{Two-fluid model}
\label{sec:model}

\subsection{Model description}

\begin{figure}[tbh]
\begin{center}
\includegraphics[scale=0.35]{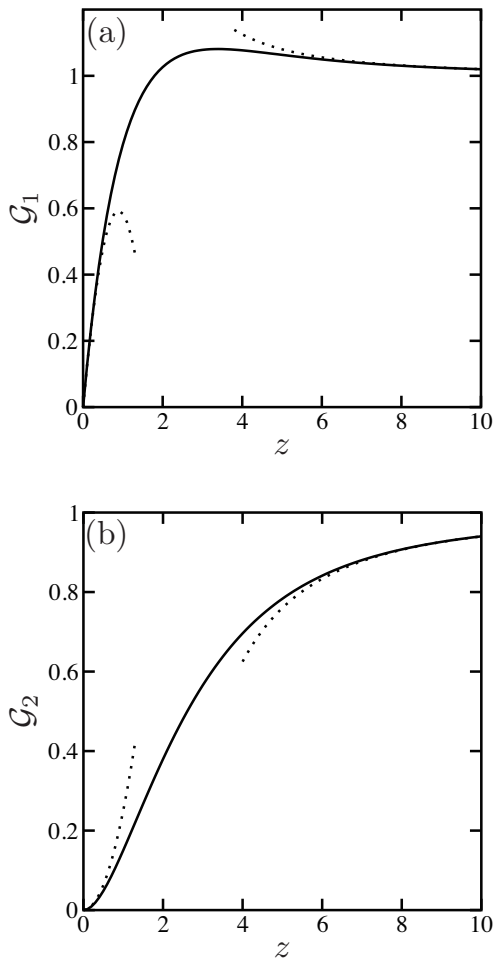}
\end{center}
\caption{
Scaling functions (a) $\mathcal{G}_1$ and (b) $\mathcal{G}_2$ [see Eq.~(\ref{G1(x)}) and 
(\ref{G2(x)}), respectively] appearing in the partial Green's function of the two-fluid model.
The scaling variable is $z=r/\xi$, where $r$ is the distance and $\xi$ is the frequency-dependent 
characteristic length scale [see Eq.~(\ref{xiomega})].
The asymptotic behaviors of these scaling functions, as analytically given by 
Eqs.~(\ref{G1(x)asymp}) and (\ref{G2(x)asymp}), respectively, are plotted with dotted lines. 
}
\label{g1g2}
\end{figure}

To describe viscoelastic media from a general point of view, we employ the two-fluid 
model that has been broadly used to describe the dynamics of polymer 
gels~\cite{deGennes76a,deGennes76b,Brochard77,Diamant15}. 
As schematically shown in Fig.~\ref{two-fluid}, there are two dynamical fields in this 
model; the displacement field $\mathbf{u}(\mathbf{r},t)$ 
of the elastic network and the velocity field $\mathbf{v}(\mathbf{r},t)$ of the permeating fluid.
Here $\mathbf{r}$ is the 3D position vector and $t$ is the time. 
The coupled dynamical equations for these two field variables are given by 
\begin{align}
\rho_u\frac{\partial^2 \mathbf{u}}{\partial t^2}
&=\mu \nabla^2 \mathbf{u}+(\mu+\lambda)\nabla(\nabla\cdot \mathbf{u}) \nonumber \\
&-\Gamma \left(\frac{\partial \mathbf{u}}{\partial t} -\mathbf{v} \right)+\mathbf{f}_u,
\label{model1}
\end{align}
\begin{align}
\rho_v\frac{\partial \mathbf{v}}{\partial t}=
\eta \nabla^2 \mathbf{v} -\nabla p 
-\Gamma \left(\mathbf{v}- \frac{\partial \mathbf{u}}{\partial t}\right)+\mathbf{f}_v.
\label{model2}
\end{align}
In the above, $\rho_u$ and $\rho_v$ are the mass densities of the two components, 
$\mu$ and $\lambda$ are the Lam\'e coefficients of the elastic network, respectively,
$\eta$ is the shear viscosity of the fluid,  $p(\mathbf{r},t)$ is the pressure field, while 
$\mathbf{f}_u$ and $\mathbf{f}_v$ are external force densities.
The elastic and the fluid components are coupled through the mutual friction characterized 
by the friction coefficient $\Gamma$. 
We note that $\mu$, $\lambda$, $\eta$ and $\Gamma$ are constants and do not depend 
on frequency.
When the volume fraction of the network is denoted by $\phi$, the above equations are
further supplemented by the condition of the total volume conservation 
\begin{align}
\nabla \cdot \left[ \phi \frac{\partial \mathbf{u}}{\partial t} 
+(1-\phi) \mathbf{v} \right]=0.
\label{genincompress}
\end{align}

In the following, we employ several simplifications of the model.
(i) We neglect inertial effects, which is justified at sufficiently low frequencies. 
Hence the l.h.s.\ of Eqs.~(\ref{model1}) and (\ref{model2}) are both neglected.
(ii) We assume that the volume fraction of the network is vanishingly small, i.e.,
$\phi \ll 1$.  
In this limit, Eq.~(\ref{genincompress}) can be approximated as 
\begin{align}
\nabla \cdot \mathbf{v} \approx 0.
\label{incompress}
\end{align}
This equation can be regarded as the incompressibility condition of the fluid component.

\subsection{Partial Green's function}

The above linearized equations can be solved by performing the Fourier transform in space 
and the Laplace transform in time for any function $f(\mathbf{r},t)$ as defined  by 
\begin{align}
f[\mathbf{q},s] =\int_{-\infty}^{\infty} {\rm d}^3 r 
\int_{0}^{\infty} {\rm d} t  \, f(\mathbf{r}, t)e^{-i\mathbf{q} \cdot \mathbf{r}-s t}.
\label{fourierlaplace} 
\end{align}
Here $\mathbf{q}$  is the 3D wavevector and $s$ the frequency in the Laplace domain. 
The general Green's function (represented by a $6\times6$ matrix) connecting 
$\mathbf{u}$ and $\mathbf{v}$ to $\mathbf{f}_u$ and $\mathbf{f}_v$ was calculated 
by Levine and Lubensky~\cite{Levine00,Levine01}.

In this paper, we particularly focus on the response of the fluid velocity $\mathbf{v}$ 
due to the point force $\mathbf{f}_v$, and use the partial Green's function defined by 
\begin{align}
v_\alpha[\mathbf{q}, s]=G_{\alpha\beta}[\mathbf{q},s] 
f_{v,\beta}[\mathbf{q},s].
\label{eq:linear}
\end{align}
Hereafter, the Einstein summation convention over repeated indices is employed.
According to Refs.~\cite{Levine00,Levine01,Sonn14,Sonn14-2}, the Green's function is given by 
\begin{align}
G_{\alpha\beta}[\mathbf{q},s] =
\frac{1+(\eta_{\rm b}/\eta)\xi^2q^2}{\eta_{\rm b} q^2(1+\xi^2q^2)}
\left( \delta_{\alpha \beta}- \hat{q}_\alpha \hat{q}_\beta \right),
\label{eq:oseen_q}
\end{align}
with $q=\vert \mathbf{q} \vert$ and $\hat{\mathbf{q}}= \mathbf{q}/q$
(see Appendix~\ref{appgreenfunction} for a detailed derivation).
In the above, the frequency-dependent bulk viscosity and characteristic length scale are 
defined by 
\begin{align}
\eta_{\rm b}=\eta + \frac{\mu}{s},~~~~~
\xi=\left(\frac{\mu\eta}{s\Gamma\eta_{\rm b}}\right)^{1/2},
\label{xiomega}
\end{align}
respectively. 
Notice that the above $3\times 3$ matrix is nothing but the part of the general $6\times 6$
matrix~\cite{Levine01}. 
In order to study the effects of molecular motors that generate forces in the cytoskeleton, one 
needs to take into account $\mathbf{f}_u$ as discussed in Refs.~\cite{Mac08,Levine09} 
and recently by us~\cite{Yasuda17}.
We note here that the partial Green's function in Eq.~(\ref{eq:oseen_q}) does not depend on 
the compressional Lam\'e coefficient $\lambda$, while it appears in the general $6\times 6$ matrix.

The Green's function in Eq.~(\ref{eq:oseen_q}) can be inverted back from 
the Fourier space to the real space (but remaining in the Laplace domain).
Following the calculation in the Appendix~\ref{appgreenfunction}, we obtain 
\begin{align}
G_{\alpha\beta}[\mathbf{r},s] & =  \frac{1}{8\pi\eta r}
\biggl[ 
\left(1+\frac{1-\eta_{\rm b}/\eta}{\eta_{\rm b}/\eta} \mathcal{G}_1(r/\xi)\right)
\delta_{\alpha \beta}
\nonumber \\
& +\left(1+\frac{1-\eta_{\rm b}/\eta}{\eta_{\rm b}/\eta}\mathcal{G}_2(r/\xi)\right)
\hat{r}_{\alpha} \hat{r}_{\beta}
\biggr].
\label{fullgreen}
\end{align}
where $r=\vert \mathbf{r} \vert$ and $\hat{\mathbf{r}}= \mathbf{r}/r$.
Here we have defined the two scaling functions by 
\begin{align}
\mathcal{G}_1(z)=1+\frac{2}{z^2}-2e^{-z}
\left(1+\frac{1}{z}+\frac{1}{z^2}\right),
\label{G1(x)}
\end{align}
\begin{align}
\mathcal{G}_2(z)=1-\frac{6}{z^2}+2e^{-z}\left(1+\frac{3}{z}+\frac{3}{z^2}\right).
\label{G2(x)}
\end{align}
In Fig.~\ref{g1g2}, we plot both $\mathcal{G}_1(z)$ and $\mathcal{G}_2(z)$ as a function 
of $z=r/\xi$. 
When $\mathcal{G}_1(z)=\mathcal{G}_2(z)=0$, the Green's function $G_{\alpha\beta}$
is purely determined by $\eta$, whereas it is fully described by $\eta_{\rm b}$ when 
$\mathcal{G}_1(z)=\mathcal{G}_2(z)=1$.

\subsection{Asymptotic expressions}

Next we discuss the asymptotic behaviors of the partial Green's function and the scaling functions.
We first note that the asymptotic expressions of the scaling functions are given by 
\begin{align}
\mathcal{G}_1(z)\approx 
\begin{cases}
4 z/3-3 z^2/4, & z \ll 1,\\
1+2/z^2, & z \gg 1,
\end{cases}
\label{G1(x)asymp}
\end{align}
\begin{align}
\mathcal{G}_2(z)\approx 
\begin{cases}
z^2/4, & z \ll 1,\\
1-6/z^2, & z \gg  1.
\end{cases}
\label{G2(x)asymp}
\end{align}
These asymptotic behaviors are also plotted in Fig.~\ref{g1g2} by the dotted lines which provide
a good approximation especially for $z \gg 1$.

For our later purpose, we focus here on the large scale behavior of the Green's function. 
For $r \gg \xi$, we obtain  
\begin{align}
G_{\alpha\beta}[\bm r,s] & \approx 
\frac{1}{8\pi\eta r}\frac{s\tau}{1+s\tau}
(\delta_{\alpha \beta}+\hat{r}_{\alpha} \hat{r}_{\beta})
\nonumber \\
& -\frac{\ell^2}{4\pi\eta r^3}\frac{1}{(1+s\tau)^2}
(\delta_{\alpha \beta}-3\hat{r}_{\alpha} \hat{r}_{\beta}),
\label{eq:Glong}
\end{align}
where we have introduced the characteristic length and time scales as
\begin{align}
\ell=(\eta/\Gamma)^{1/2},~~~~~
\tau=\eta/\mu.
\label{defell}
\end{align}
As argued in Ref.~\cite{Sonn14},  the first and the second terms of Eq.~(\ref{eq:Glong}) 
are proportional to $1/r$ and $\ell^2/r^3$, respectively. 
The competition between these two terms is characterized by the crossover length $\ell$.

\subsection{Coupling mobilities}

In the following sections, we shall consider correlated motions of two point particles
embedded in the fluid component.
For this purpose, we shall introduce the coupling mobility between the two points
$M_{\alpha \beta}[r,s]$ that is directly related to the partial Green's function in  
Eq.~(\ref{fullgreen}).
Since $G_{\alpha \beta}$ is generally expressed as 
$G_{\alpha \beta}=C_1\delta_{\alpha \beta} + C_2\hat{r}_\alpha \hat{r}_\beta$,  
the ``longitudinal" and the ``transverse" coupling mobilities are given by 
$M_{xx}=C_1+C_2$ and $M_{yy}=C_1$, respectively. 
Hence they are
\begin{align}
M_{xx}[r,s]=\frac{1}{4\pi\eta r}
\left[1-\frac{\mathcal{G}_1(z)+\mathcal{G}_2(z)}{2(1+s \tau)}\right],
\label{longitudinal}
\end{align}
\begin{align}
M_{yy}[r,s]=\frac{1}{8\pi\eta r}
\left[1-\frac{\mathcal{G}_1(z)}{1+s \tau}\right],
\label{transverse}
\end{align}
where $z=r/\xi=(r/\ell)\sqrt{1+s \tau}$.
We shall use these coupling mobilities in order to calculate various correlation functions 
in the next sections. 
Since $M_{xy}=0$ by symmetry, it is sufficient to consider only the above two coupling 
mobilities~\cite{Sonn14}.

\section{Passive two-point correlation functions}
\label{sec:pass2pcorr}

Here we discuss the correlated dynamics of two distinctive passive particles 
immersed in a viscoelastic gel that is in thermal equilibrium.
This situation is relevant to the ``two-point microrheology" experiments as discussed 
before~\cite{CVWGKYW}. 
Compared to the ``single-particle microrheology" (with the use of a finite size particle), there 
are several advantages to perform multi-particle microrheology~\cite{SM10}.

\subsection{Fluctuation dissipation theorem}

Consider a pair of point particles undergoing Brownian motion separated by a distance $d$ 
as shown in Fig.~\ref{two-fluid} (but without force dipoles).
We denote the positions of these two point particles by 
$\mathbf{R}_1(t)=\mathbf{R}_1 + \Delta \mathbf{R}_1(t)$ and 
$\mathbf{R}_2(t)=\mathbf{R}_2 + \Delta \mathbf{R}_2(t)$, where 
$d= \vert \mathbf{R}_{2}-\mathbf{R}_{1} \vert$. 
Then the velocities of these point particles are given by 
$\mathbf{V}_1(\mathbf{R}_1,t) = \Delta \dot{\mathbf{R}}_1(t)$ and 
$\mathbf{V}_2(\mathbf{R}_2,t) = \Delta \dot{\mathbf{R}}_2(t)$.
The quantities of interest are the velocity cross-correlation function (CCF) 
$\langle V_{1\alpha} V_{2 \alpha'}(t)\rangle_d$, and the displacement 
CCF $\langle \Delta R_{1\alpha} \Delta R_{2 \alpha'}(t)\rangle_d$.
Without loss of generality, we define the $x$-axis to be along the line connecting 
the two particles, i.e., $\mathbf{R}_{2}-\mathbf{R}_{1}=d \hat{\mathbf{e}}_x$.

According to the fluctuation dissipation theorem (FDT), the velocity CCFs in thermal 
equilibrium are related to the coupling mobility in the Laplace domain by~\cite{SM10,Sonn14}
\begin{equation}
\langle V_{1\alpha} V_{2\alpha'}[s]\rangle_d = k_{\rm B}T M_{\alpha \alpha'}[r=d,s],
\label{2pvelvel}
\end{equation}
where $k_{\rm B}$ is the Boltzmann constant, $T$ the temperature. 
The power spectral density (PSD) of the passive velocity CCF can be obtained by using the relation 
\begin{equation}
\langle V_{1\alpha}V_{2 \alpha'} (\omega)\rangle_d 
=2\Re \langle V_{1\alpha}V_{2 \alpha'}[s=i\omega]\rangle_d, 
\label{passivepsd}
\end{equation}
where $\omega$ is the frequency in the Fourier domain, and $\Re$ indicates the real part.

\begin{figure}[tbh]
\begin{center}
\includegraphics[scale=0.35]{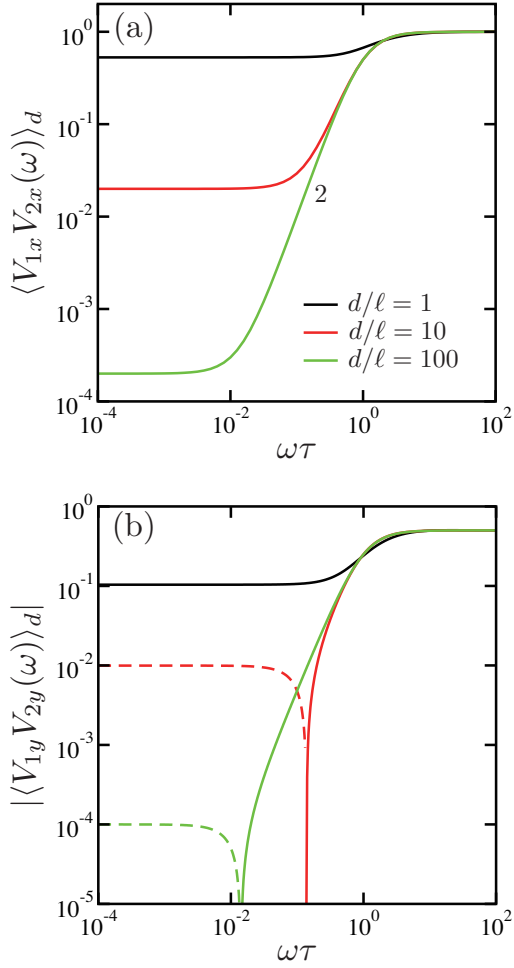}
\end{center}
\caption{
The passive component of the power spectral density (PSD) of the two-point velocity 
cross-correlation functions (CCFs)  
(a) $\langle V_{1x}V_{2x}(\omega)\rangle_d$ and 
(b) $\vert \langle V_{1y}V_{2y}(\omega)\rangle_d \vert$ 
[see Eq.~(\ref{passivepsd})] 
as a function of $\omega \tau$ for $d/\ell=1$, $10$, $100$.  
Here $\tau=\eta/\mu$ is the viscoelastic time scale, and $d$ is the distance 
between the two point particles immersed in viscoelastic media described by the two-fluid model. 
Both CCFs are scaled by $k_{\rm B}T/(2\pi\eta d)$ in order to make 
them dimensionless. 
Since $\langle V_{1y}V_{2y}(\omega)\rangle_d$ takes negative values for smaller 
$\omega \tau$  (shown by the dashed lines), we have plotted in (b) its absolute value.
The number ``2" in (a) indicates the slope representing the exponent of the power-law
behaviors.
}
\label{PSDxxyy}
\end{figure}

\begin{figure}[tbh]
\begin{center}
\includegraphics[scale=0.35]{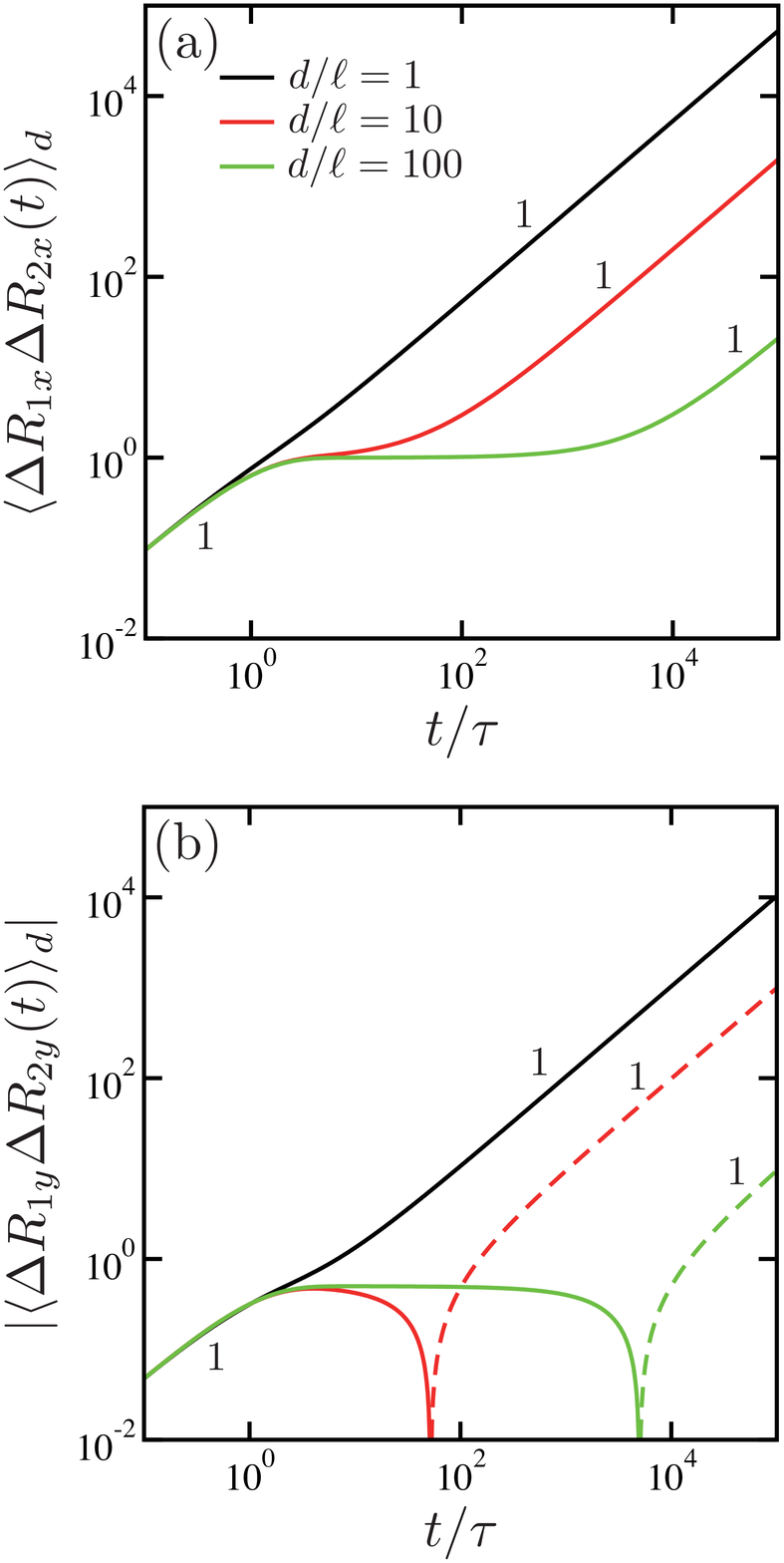}
\end{center}
\caption{
The passive component of the two-point displacement cross-correlation functions (CCFs) 
(a) $\langle \Delta R_{1x} \Delta R_{2x}(t)\rangle_d$ and 
(b) $\vert \langle \Delta R_{1y} \Delta R_{2y}(t)\rangle_d \vert$ 
[see Eq.~(\ref{msdtime})] as a function of $t/\tau$ for $d/\ell=1$, $10$, $100$.
Here $d$ is the distance between the two point particles immersed in viscoelastic media.
Both CCFs are scaled by $k_{\rm B}T \tau/(2\pi \eta d)$ in order to make them 
dimensionless. 
Since $\langle \Delta R_{1y} \Delta R_{2y}(t)\rangle_d$ takes negative values 
for larger $t/ \tau$ (shown by the dashed lines), we have plotted in (b) its absolute value.
The numbers indicate the slope representing the exponent of the power-law behaviors.
}
\label{2pmsd}
\end{figure}

In Fig.~\ref{PSDxxyy}, we plot the scaled PSDs $\langle V_{1x}V_{2x}(\omega)\rangle_d$
and $\langle V_{1y}V_{2y}(\omega)\rangle_d$ as a function of $\omega \tau$
using the longitudinal and the transverse coupling mobilities obtained in Eqs.~(\ref{longitudinal}) 
and (\ref{transverse}), respectively.
Different colors represent different distances, $d$, between the two points.
In Fig.~\ref{PSDxxyy}(a), $\langle V_{1x}V_{2x}(\omega)\rangle_d$ increases 
for $\omega \tau > \sqrt{2}(\ell/d)$ as $\sim \omega^2$ [see later Eqs.~(\ref{passvxvx}) 
and (\ref{passvyvy})], while it saturates for $\omega \tau > 1$. 
Since $\langle V_{1y}V_{2y}(\omega)\rangle_d$ takes negative values for smaller 
$\omega \tau$ when $d/\ell=10$ and $100$, we have plotted its absolute value in 
Fig.~\ref{PSDxxyy}(b). 
Notice that in both (a) and (b), the PSDs are scaled by $k_{\rm B}T/(2\pi\eta d)$, and 
take the same asymptotic value in the large $\omega \tau$ limit.

The passive displacement CCF in thermal equilibrium as a function of time can be directly 
obtained by the following inverse Laplace transform of the velocity CCF~\cite{Mason00,SM10}: 
\begin{equation}
\langle \Delta R_{1\alpha} \Delta R_{2 \alpha'}(t)\rangle_d=
\frac{1}{2 \pi i}
\int_{c-i\infty}^{c+i\infty} {\rm d}s \, 
\frac{2}{s^2} 
\langle V_{1\alpha} V_{2\alpha'}[s] \rangle_d e^{st},
\label{msdtime}
\end{equation}
where $c$ is a real number. 
Performing the numerical inverse Laplace transform of Eq.~(\ref{msdtime}), we plot in 
Fig.~\ref{2pmsd} the longitudinal and the transverse CCFs 
$\langle \Delta R_{1x} \Delta R_{2x}(t)\rangle_d$ and 
$\langle \Delta R_{1y} \Delta R_{2y}(t)\rangle_d$ 
as a function of $t/\tau$  for different distances $d$ between the two points.

In Fig.~\ref{2pmsd}(a), the longitudinal CCF is proportional to $t$ for $t/\tau < 1$ 
and $t/\tau > (d/\ell)^2/2$
[see later Eq.~(\ref{pass2pmsdxx})]. 
In the former time region which is smaller than the viscoelastic time scale $\tau=\eta/\mu$, 
the two point particles interact through the fluid component of the two-fluid model. 
In the latter long time region, on the other hand, the CCF obeys the normal diffusive 
behavior as expected for any viscoelastic material with a characteristic relaxation time. 
Between these crossover time scales, the CCF remains almost constant due to the 
elastic component that suppresses the motion of the point particles.  
This is because the elastic property of the medium, representing the polymer network, 
is pronounced in these time scales. 
For $d/\ell=1$, on the other hand, the CCF is almost proportional to $t$ during the entire 
time region.

In Fig.~\ref{2pmsd}(b), the absolute value of 
$\langle \Delta R_{1y} \Delta R_{2y}(t)\rangle_d$ is plotted because it takes 
negative values for larger $t$. 
This means that the relative transverse motion of the two point particles is 
anti-correlated when their separation $d/\ell$ is large enough.
Nevertheless, the general time-dependent behavior of the transverse CCF is almost the 
same as that of the longitudinal one in (a).

The crossover behaviors of the passive displacement CCF for large $d/\ell$ showing the 
successive scaling as $t \rightarrow t^0 \rightarrow t$ can explain some of the apparent 
power-law behaviors of soft matter~\cite{Chen10} or biological cells~\cite{Hoffman09}. 
It should be noted, however, that the passive displacement CCF in thermal equilibrium 
exhibits only a sub-diffusive behavior.

\subsection{Large distance behaviors}

In the limit of large distances $d \gg \ell$ between the two points, we can use Eq.~(\ref{eq:Glong})
for the partial Green's function to obtain the PSDs in the Fourier domain as 
\begin{align}
&\langle V_{1x}V_{2x}(\omega)\rangle_d \approx \frac{k_{\rm B}T}{2\pi\eta d}
\frac{(\omega\tau)^2}{1+(\omega \tau)^2}
\nonumber \\
& +\frac{k_{\rm B}T\ell^2}{\pi\eta d^3}
\left[\frac{1}{[1+(\omega\tau)^2]^2}-
\frac{(\omega\tau)^2}{[1+(\omega\tau)^2]^2}\right],
\label{passvxvx}
\end{align}
\begin{align}
&\langle V_{1y}V_{2y}(\omega)\rangle_d \approx \frac{k_{\rm B}T}{4\pi\eta d}
\frac{(\omega\tau)^2}{1+(\omega\tau)^2}
\nonumber \\
& -\frac{k_{\rm B}T\ell^2}{2\pi\eta d^3}
\left[\frac{1}{[1+(\omega\tau)^2]^2}-
\frac{(\omega\tau)^2}{[1+(\omega\tau)^2]^2}\right].
\label{passvyvy}
\end{align}

For the large distance behavior of the displacement CCFs,  we obtain 
\begin{align}
& \langle \Delta R_{1x} \Delta R_{2x} (t) \rangle_d \approx
\frac{k_{\rm B}T\tau}{2\pi\eta d}(1-e^{-t/\tau})
\nonumber \\
&+\frac{k_{\rm B}T\tau\ell^2}{\pi\eta d^3}
\left[\frac{t}{\tau}(1+e^{-t/\tau})-2(1-e^{-t/\tau})\right],
\label{pass2pmsdxx}
\end{align}
\begin{align}
&\langle \Delta R_{1y} \Delta R_{2y} (t) \rangle_d \approx
\frac{k_{\rm B}T\tau}{4\pi\eta d}(1-e^{-t/\tau})
\nonumber \\
&-\frac{k_{\rm B}T\tau\ell^2}{2\pi\eta d^3}
\left[\frac{t}{\tau}(1+e^{-t/\tau})-2(1-e^{-t/\tau})\right].
\label{pass2pmsdyy}
\end{align}
In the above expressions, the first term is proportional to $t$ in the short time regime, 
whereas it saturates in the long time limit. 
Whereas the second term in each expression is proportional to $t$ in the 
long time limit, which dominates the large scale behavior. 
These properties of the displacement CCF can be clearly observed in Fig.~\ref{2pmsd} 
especially for larger $d/\ell$.
Although not plotted, Eqs.~(\ref{pass2pmsdxx}) and (\ref{pass2pmsdyy}) almost 
completely recover the numerical plots in Fig.~\ref{2pmsd}.

\section{Active one-point correlation functions}
\label{sec:1pcorr}

In this section, we shall consider the collective advection effects due to active force dipoles 
on passive particles immersed in viscoelastic media.
A simple ``dimer model" for a stochastic hydrodynamic force dipole was previously 
discussed in Refs.~\cite{Mikhailov15,Kapral16}. 
To investigate the hydrodynamic effects of the force dipoles, we employ the partial Green's function 
representing the response of the fluid velocity $\mathbf{v}$ due to the force $\mathbf{f}_v$ 
acting on the fluid component [see Eq.~(\ref{eq:linear})]. 
This is different from Refs.~\cite{Mac08,Levine09} where they discussed the effects of molecular 
motors generating forces in the cytoskeleton which corresponds to the elastic component of
the two-fluid model. 
Our aim is to focus on the role of active force dipoles that exist in the fluid component.
A more unified treatment of these two different sources of active forces has been investigated 
in our separate publication~\cite{Yasuda17}.

\subsection{Velocity induced by active force dipoles}

When a point force $\mathbf{f}_v$ is applied to the fluid at a point $\mathbf{r}$, 
it induces a fluid velocity at another position $\mathbf{R}$ that advects a point 
particle located there.
As in the previous section, we denote the position of this passive point particle by 
$\mathbf{R}(t)=\mathbf{R}_0 + \Delta \mathbf{R}(t)$, and its velocity by
$\mathbf{V}(\mathbf{R},t) = \Delta \dot{\mathbf{R}}(t)$.
Using the Green's function calculated in Sec.~\ref{sec:model}, we obtain the relation
between $\mathbf{V}$ and $\mathbf{f}_v$ as
\begin{align}
V_{\alpha}(\mathbf{R},t)=\int_{-\infty}^t {\rm d}t'\,
G_{\alpha\beta}(\mathbf{R}-\mathbf{r},t-t') 
f_{v,\beta}(\mathbf{r},t').
\end{align}

Consider an oscillating dimer of length $a(t)$ and the force magnitude $f_{\rm d}(t)$ 
with its orientation given by the unit vector $\hat{\mathbf{e}}$.
In this case, the induced velocity of a passive particle at $\mathbf{R}$ due to the  oscillating 
dimer is given by~\cite{Mikhailov15} 
\begin{align}
V_\alpha(\mathbf{R},t) \approx\int_{-\infty}^t {\rm d}t'\,
\frac{\partial G_{\alpha\beta}(\mathbf{R}-\mathbf{r},t-t')}
{\partial r_\gamma}\hat{e}_\beta \hat{e}_\gamma m(t'),
\label{valphart}
\end{align}
where we have used the approximation $a \ll \vert \mathbf{R}-\mathbf{r} \vert$, and 
$m(t)=a(t)f_{\rm d}(t)$ denotes the magnitude of the force dipole.

We further consider a collection of such active force dipoles, located at positions
$\{ \mathbf{R}_i \}$ with orientations $\{ \hat{\mathbf{e}}_i \}$.
By summing up for all the dipoles, the velocity of the passive particle is then given by~\cite{Mikhailov15} 
\begin{align}
V_\alpha(\mathbf{R}_0,t) & \approx \int_{-\infty}^t {\rm d}t'\, \int {\rm d}^3r\, 
\frac{\partial G_{\alpha\beta}(\mathbf{r},t-t')}{\partial r_\gamma}
\nonumber \\
& \times \sum_i \hat{e}_{i,\gamma } \hat{e}_{i,\beta} m_i(t') 
\delta(\mathbf{R}_i-\mathbf{R}_0-\mathbf{r}),
\label{acivevelocity}
\end{align}
where we have assumed that the displacement of the passive particle is small, and kept
only the lowest order term. 
This equation describes the motion of a passive point particle due to non-thermal active 
noise arising from the collective operation of active force dipoles.

Hereafter we introduce the bilateral Fourier transform in time for any function $f(t)$ as 
\begin{align}
f(\omega) =\int_{-\infty}^{\infty} {\rm d} t  \, f(t) e^{-i \omega t}, 
\end{align}
[cf.\ Eq.~(\ref{fourierlaplace})].
Performing the bilateral Fourier transform of Eq.~(\ref{acivevelocity}), we obtain  
\begin{align}
V_\alpha(\mathbf{R}_0,\omega) &\approx \int {\rm d}^3r\, 
\frac{\partial G_{\alpha\beta}[\mathbf{r},\omega]}{\partial r_\gamma}
\nonumber \\
&\times \sum_i \hat{e}_{i,\gamma }\hat{e}_{i,\beta} m_i(\omega) \delta(\mathbf{R}_i-\mathbf{R}_0-\mathbf{r}),
\label{vomega}
\end{align}
where $G_{\alpha\beta}[\mathbf{r},\omega]=G_{\alpha\beta}[\mathbf{r},s=i\omega]$.
We shall use this expression to calculate the velocity correlation functions and the 
mean squared displacement (MSD) of the passive particle.

\subsection{Active auto-correlation functions}

We now calculate the velocity auto-correlation function (ACF) of a passive particle located 
on average at $\mathbf{R}_0$. 
If the correlation between different force dipoles vanishes, i.e., 
$\langle m_i m_j(\omega) \rangle =\langle m^2(\omega) \rangle \delta_{ij}$, 
we get from Eq.~(\ref{vomega})
\begin{align}
& \langle V_\alpha V_{\alpha'}(\mathbf{R}_0, \omega)\rangle
\nonumber \\
& = \int {\rm d}^3r\, \frac{\partial G_{\alpha\beta}[\mathbf{r},\omega]}{\partial r_\gamma}
\frac{\partial G_{\alpha'\beta'}[\mathbf{r},-\omega]}{\partial r_{\gamma'}}
\nonumber \\
&\times \sum_i \langle \hat{e}_{i,\gamma }\hat{e}_{i,\beta}\hat{e}_{i,\gamma' }\hat{e}_{i,\beta'}\rangle 
\langle m_i^2(\omega)\rangle 
\langle \delta(\mathbf{R}_i-\mathbf{R}_0-\mathbf{r})\rangle 
\nonumber \\
&=\Omega_{\beta\beta'\gamma\gamma'} \langle m^2(\omega)\rangle
\nonumber \\
& \times \int {\rm d}^3r\, \frac{\partial G_{\alpha\beta}[\mathbf{r},\omega]}{\partial r_\gamma}
\frac{\partial G_{\alpha'\beta'}[\mathbf{r},-\omega]}{\partial r_{\gamma'}} 
c(\mathbf{R}_0 + \mathbf{r}),
\end{align}
where $c(\mathbf{r})=\sum_i  \langle \delta(\mathbf{R}_i-\mathbf{r})\rangle$ is the
local concentration of force dipoles at a point $\mathbf{r}$ in the fluid component.
In the above, a symbol
\begin{align}
\Omega_{\beta\beta'\gamma\gamma'} 
& =\langle \hat{e}_{\beta} \hat{e}_{\beta'} \hat{e}_{\gamma} \hat{e}_{\gamma' }\rangle
\nonumber \\
& =\frac{1}{15}(\delta_{\beta\beta'}\delta_{\gamma\gamma'}
+\delta_{\beta\gamma}\delta_{\beta'\gamma'}
+\delta_{\beta\gamma'}\delta_{\beta'\gamma}),
\end{align}
has been defined, and we have assumed that the orientations of active force dipoles are
not correlated with their positions. 
In other words, we do not consider any nematic ordering of force dipoles~\cite{Mikhailov15}.

\begin{figure}[tbh]
\begin{center}
\includegraphics[scale=0.35]{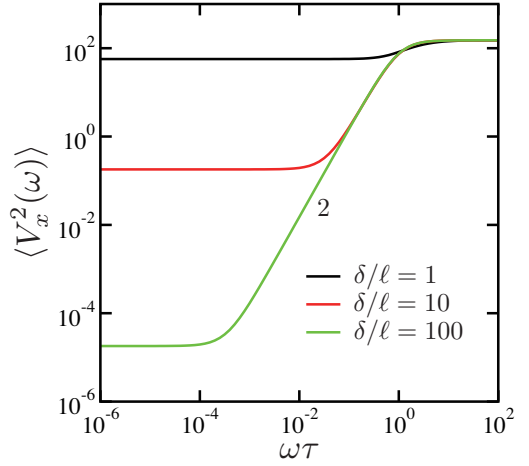}
\end{center}
\caption{
The active component of the power spectral density (PSD) $\langle V_x^2(\omega)\rangle$
[see Eq.~(\ref{iomega})] as a function of $\omega \tau$ for $\delta/\ell=1$, $10$, $100$.
Here a single point particle is immersed in viscoelastic media described by the two-fluid model, 
and $\delta$ is the cutoff length corresponding to the particle size.
In the plot, the PSD is scaled by $c_0\langle m^2(\omega)\rangle/(2880\pi^2\eta^2\delta)$ 
in order to make it dimensionless.
The number indicates the slope representing the exponent of the power-law behavior.
}
\label{1pointPSD}
\end{figure}

When active force dipoles are uniformly distributed in space with a constant concentration, 
$c(\mathbf{r})=c_0$, the velocity ACF $\langle V_\alpha V_{\alpha'}(\omega)\rangle$
is isotropic, i.e.,  $\langle V_x^2(\omega)\rangle=\langle V_y^2(\omega)\rangle
=\langle V_z^2(\omega)\rangle$ and vanishes otherwise.  
Hence it is enough to consider only the $x$-direction, and we obtain the active PSD  as
\begin{align}
\langle V_x^2(\omega)\rangle & =  \frac{c_0}{3}
\Omega_{\beta\beta'\gamma\gamma'} \langle m^2 (\omega)\rangle
\nonumber \\
& \times \int {\rm d}^3r\, \frac{\partial G_{\alpha\beta}[\mathbf{r},\omega]}{\partial r_\gamma}
\frac{\partial G_{\alpha\beta'}[\mathbf{r},-\omega]}{\partial r_{\gamma'}}
\nonumber \\
& =  \frac{1}{3\cdot 8^2 \cdot 15 \pi^2}\frac{c_0}{\eta^2\ell}
\langle m^2 (\omega)\rangle\mathcal{I}(\omega).
\label{iomega}
\end{align}
Here we have introduced the scaled PSD defined by
\begin{align}
\mathcal{I}(\omega)=15\Omega_{\beta\beta'\gamma\gamma'}
\int {\rm d}^3\bar{r}\, \frac{\partial g_{\alpha\beta}[\mathbf{r},\omega]}{\partial \bar{r}_\gamma}
\frac{\partial g_{\alpha\beta'}[\mathbf{r},-\omega]}{\partial \bar{r}_{\gamma'}},
\label{Iomega}
\end{align}
together with 
$g_{\alpha\beta}[\mathbf{r},\omega]=
8\pi\eta\ell G_{\alpha\beta}[\mathbf{r},s=i\omega]$ and $\bar{r}=r/\ell$.
Since the above integral diverges for short length scales, we need to introduce a small 
cutoff length $\delta$. 
Physically, $\delta$ can be regarded as the size of the passive particle.
In Fig.~\ref{1pointPSD}, we numerically plot the scaled PSD, $\mathcal{I}(\omega)$,
as a function of $\omega \tau$ for different cutoff lengths $\delta/\ell=1$, $10$ and $100$. 
These values correspond to the situation when the passive particle is larger than the mesh size.
For $\omega \tau > 2\sqrt{3}(\ell/\delta)^2$, the PSD increases as $\sim \omega^2$ and 
saturates for $\omega \tau > 1$ [see later Eq.~(\ref{eq:spectol_high})]. 
Note that the asymptotic value of the scaled PSD for large $\omega \tau$ is independent of $\delta$.

If we use Eq.~(\ref{eq:Glong}) for the partial Green's function $G_{\alpha\beta}$ in the large distance
limit $r\gg \ell$, the scaled PSD can be approximately calculated as  
\begin{align}
\mathcal{I}(\omega) & \approx 
48 \pi \left[ \frac{(\omega\tau)^2}{1+ (\omega \tau)^2} 
\left( \frac{\ell}{\delta} \right) 
-\frac{4(\omega\tau)^2}{[1+ (\omega \tau)^2]^2}\left( \frac{\ell}{\delta} \right)^3
\right. \nonumber \\
& \left. +\frac{12}{[1+ (\omega \tau)^2]^2}\left( \frac{\ell}{\delta} \right)^5 \right].
\label{eq:spectol_high}
\end{align}
The detailed derivation of this expression is given in the Appendix~\ref{appderivepsd}.
Although not plotted, we have confirmed that Eq.~(\ref{eq:spectol_high}) perfectly reproduces 
the curve of $\delta/\ell=100$ in Fig.~\ref{1pointPSD}.
We should keep in mind, however, that to regard the cutoff length $\delta$ as the 
particle size is only an approximation, and hence the numerical prefactor should not be
taken as accurate when we compare with experiments.

\subsection{Uncorrelated force dipoles}

\begin{figure}[tbh]
\begin{center}
\includegraphics[scale=0.35]{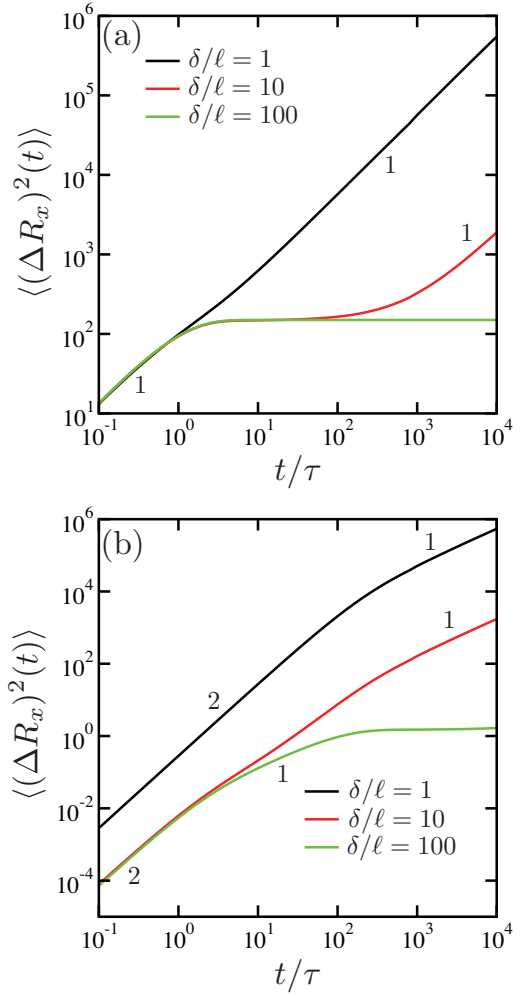}
\end{center}
\caption{
The active component of the mean squared displacement (MSD) $\langle (\Delta R_x)^2(t)\rangle$ 
[see Eq.~(\ref{msdtimefourier})] as a function of $t/ \tau$ for $\delta/\ell=1$, $10$, $100$.
Here a single point particle is immersed in viscoelastic media, and $\delta$ is the cutoff length.
(a) The case when the time correlation of the force dipole is $\delta$-correlated
[see Eq.~(\ref{deltacorr})]. 
(b) The case when the time correlation of the force dipole decays exponentially with a 
characteristic relaxation time $\tau_{\rm d}$ [see Eq.~(\ref{expcorr})], and we set here 
$\tau_{\rm d}/\tau=100$.
In these plots, $\langle (\Delta R_x)^2(t)\rangle$ is scaled by $c_0 S\tau/(2880 \pi^2\eta^2\delta)$
in order to make it dimensionless.
The numbers indicate the slope representing the exponent of the power-law behavior.
}
\label{1pointMSD}
\end{figure}

In order to further calculate the active PSD, the statistical property for the time correlation 
of a force dipole needs to be specified.
First we assume that it is only $\delta$-correlated in time and is given by
\begin{align}
\langle m(t)m(t')\rangle=S\delta(t-t'),
\label{deltacorr}
\end{align}
where $S$ fixes the fluctuation amplitude.
In the Fourier representation, this simply means that $\langle m^2 (\omega)\rangle=S$. 
Once we know the active PSD of the velocity ACF, the corresponding MSD of a passive 
particle in the $x$-direction can be obtained by the inverse Fourier transform:
\begin{equation}
\langle (\Delta R_x)^2(t)\rangle=
\int_{-\infty}^{\infty} 
\frac{{\rm d} \omega}{2 \pi }\,  
\frac{2}{(i\omega)^2} \langle V_x^2(\omega)\rangle e^{i\omega t}.
\label{msdtimefourier}
\end{equation}
In contrast to the inverse Laplace transform in Eq.~(\ref{msdtime}), we also take into account 
the initial condition by including a constant term in the above transformation.
In Fig.~\ref{1pointMSD}(a), we numerically plot the scaled $\langle (\Delta R_x)^2(t)\rangle$ 
as a function of $t/\tau$ for $\delta/\ell=1$, $10$ and $100$. 
Here the MSD is proportional to $t$ both for short time scales $t/\tau < 1$ and for long time 
scales $t/\tau > (\delta/\ell)^4/12$ [see later Eq.~(\ref{onepointmasdx})].
For the intermediate time range $1 < t/\tau < (\delta/\ell)^4/12$, on the other hand, the 
MSD is strongly suppressed due to the elastic component of the two-fluid model, and it remains 
almost constant.

If we use the asymptotic expression Eq.~(\ref{eq:Glong}) for the partial Green's function as before,  
the PSD can be obtained from Eqs.~(\ref{iomega}) and (\ref{eq:spectol_high}) as 
\begin{align}
&\langle V_x^2(\omega)\rangle\approx  \frac{1}{60\pi}\frac{c_0 S}{\eta^2\ell}
\left[
\frac{(\omega\tau)^2}{1+ (\omega \tau)^2} 
\left( \frac{\ell}{\delta} \right) 
\right.
\nonumber \\
& \left. 
-\frac{4(\omega\tau)^2}{[1+ (\omega \tau)^2]^2}\left( \frac{\ell}{\delta} \right)^3
+\frac{12}{[1+ (\omega \tau)^2]^2}\left( \frac{\ell}{\delta} \right)^5
\right].
\label{vx2psd}
\end{align}
Then, with the use of Eq.~(\ref{msdtimefourier}), the asymptotic MSD can be analytically obtained as
\begin{align}
\langle (\Delta R_x)^2(t) \rangle &\approx \frac{1}{60\pi}\frac{c_0 S \tau}{\eta^2\ell}
\left[(1-e^{-t/\tau}) \left( \frac{\ell}{\delta} \right)
\right.
\nonumber \\
&+ 2[e^{-t/\tau}(t/\tau+1)-1] \left( \frac{\ell}{\delta} \right)^3
\nonumber \\
& \left. + 6[e^{-t/\tau}(t/\tau+3)+2t/\tau-3] 
\left( \frac{\ell}{\delta} \right)^5 
\right].
\label{onepointmasdx}
\end{align}
The first term in the r.h.s.\ of the above equation indicates that the normal diffusion 
occurs for the short time scale $t \ll \tau$, while it saturates in the longer times.
In the long time limit, on the other hand, we can set $e^{-t/\tau} \approx 0$, and one finds
that MSD is proportional to $t$ for $t \gg (\delta/\ell)^4/12$ [see the third line of 
Eq.~(\ref{onepointmasdx})], as mentioned above.

\subsection{Exponentially correlated force dipoles}

Next we calculate the PSD and the MSD when the time correlation of a force dipole
decays exponentially with a characteristic relaxation time $\tau_{\rm d}$, i.e., 
\begin{align}
\langle m(t)m(t')\rangle=\frac{S}{2\tau_{\rm d}} e^{-|t-t'|/\tau_{\rm d}}.
\label{expcorr}
\end{align}
In this case, we have $\langle m^2 (\omega)\rangle=S/[1+(\omega\tau_{\rm d})^2]$ in the
Fourier representation.
Some justification of the above simple expression will be separately discussed in 
Sec.~\ref{sec:discussion}.
Mathematically, Eq.~(\ref{expcorr}) reduces to Eq.~(\ref{deltacorr}) in the limit of 
$\tau_{\rm d} \rightarrow 0$.  
Then the active PSD is given by 
\begin{align}
\langle V_x^2(\omega)\rangle &= \frac{1}{2880\pi^2}\frac{c_0 S}{\eta^2\ell}
\frac{1}{1+(\omega\tau_{\rm d})^2}\mathcal{I}(\omega),
\label{activePSD}
\end{align}
where $\mathcal{I}(\omega)$ was defined before in Eq.~(\ref{Iomega}).

In Fig.~\ref{1pointMSD}(b), we numerically plot the scaled $\langle (\Delta R_x)^2(t)\rangle$ 
as a function of $t/\tau$ when $\tau_{\rm d}/\tau=100$ for 
$\delta/\ell=1$, $10$ and $100$, 
i.e., the distance between the two points is larger than the mesh size.
For $\delta/\ell=1$, we find that the active MSD is proportional to $t^2$ and exhibits a 
super-diffusive behavior within the time region $t < \tau_{\rm d}$. 
For $\delta/\ell=100$, such a super-diffusive behavior is observed only up to the viscoelastic 
time scale $t/\tau <1$, and the MSD exhibits a normal diffusive behavior for $t/\tau>1$. 
The active MSD for $\delta/\ell=100$ is further suppressed for larger time scales. 
In the very long time limit, the active MSD will be again proportional to $t$~\cite{Yasuda17}.

Using the asymptotic expression Eq.~(\ref{eq:Glong}), the active PSD is now given by 
\begin{align}
&\langle V_x^2(\omega)\rangle\approx  \frac{1}{60\pi}\frac{c_0 S}{\eta^2\ell}
\frac{1}{1+(\omega\tau_{\rm d})^2}
\left[
\frac{(\omega\tau)^2}{1+ (\omega \tau)^2} 
\left( \frac{\ell}{\delta} \right) 
\right.
\nonumber \\
& \left. 
-\frac{4(\omega\tau)^2}{[1+ (\omega \tau)^2]^2}\left( \frac{\ell}{\delta} \right)^3
+\frac{12}{[1+ (\omega \tau)^2]^2}\left( \frac{\ell}{\delta} \right)^5
\right].
\label{vx2psdcorr}
\end{align}
Then the corresponding active one-point MSD can be obtained up to the lowest order in 
$\ell/\delta$ as 
\begin{align}
& \langle (\Delta R_x)^2(t) \rangle \approx  
\frac{1}{60\pi}\frac{c_0S\tau}{\eta^2\ell} 
\frac{1}{1+\tau_{\rm d}/\tau}
\nonumber \\
& \times \left[1+\frac{(\tau_{\rm d}/\tau)e^{-t/\tau_{\rm d}}}{1-\tau_{\rm d}/\tau}-
\frac{e^{-t/\tau}}{1-\tau_{\rm d}/\tau}\right] \left( \frac{\ell}{\delta} \right).
\label{msdlowestorder}
\end{align}
This equation reduces to the first line of Eq.~(\ref{onepointmasdx}) in the limit of 
$\tau_{\rm d} \rightarrow 0$. 
By Taylor expanding the above expression for small $t$, one can indeed show that the
linear term in $t$ vanishes, and the active MSD increases as $\sim t^2$. 
The full expression of the active one-point MSD including higher order terms is provided in the 
Appendix~\ref{appmsdhigherorder}.
The analytic expressions in Eqs.~(\ref{vx2psdcorr}) and (\ref{msdlowestorder}) are 
the general and important results of this paper.

\section{Active two-point correlation functions}
\label{sec:act2pcorr}

\subsection{Velocity cross-correlation functions}

In this section, we consider the active velocity CCF between the two points at $\mathbf{R}_1$
and $\mathbf{R}_2$ that are separated by a distance $d$, as shown in Fig.~\ref{two-fluid}
and also discussed in Sec.~\ref{sec:pass2pcorr}.
With the use of Eq.~(\ref{acivevelocity}), the active two-point velocity CCF can be evaluated by  
\begin{align}
&\langle V_{1\alpha} V_{2\alpha'}(\mathbf{R}_{1}, \mathbf{R}_{2}, \omega)\rangle_d 
\nonumber \\
&= \int {\rm d}^3r\,\frac{\partial G_{\alpha\beta}[\mathbf{r},\omega]}{\partial r_\gamma}
\left( \frac{\partial G_{\alpha'\beta'}[\mathbf{r}',-\omega]}{\partial r'_{\gamma'}}
\right)_{\mathbf{r}'=\mathbf{r}-(\mathbf{R}_2-\mathbf{R}_1)}
\nonumber \\
&\times\sum_i \langle \hat{e}_{i,\gamma }\hat{e}_{i,\beta}\hat{e}_{i,\gamma' }\hat{e}_{i,\beta'}\rangle 
\langle m_i^2(\omega)\rangle \langle 
\delta(\mathbf{R}_i-\mathbf{R}_1-\mathbf{r})\rangle 
\nonumber \\
&=\Omega_{\beta\beta'\gamma\gamma'} \langle m^2 (\omega)\rangle
\nonumber \\
& \times \int {\rm d}^3r\, \frac{\partial G_{\alpha\beta}[\mathbf{r},\omega]}{\partial r_\gamma}
\left( \frac{\partial G_{\alpha'\beta'}[\mathbf{r}',-\omega]}{\partial r'_{\gamma'}}
\right)_{\mathbf{r}'=\mathbf{r}-(\mathbf{R}_2-\mathbf{R}_1)} 
\nonumber \\ 
& \times c(\mathbf{R}_1 + \mathbf{r}). 
\end{align}

\begin{figure}[tbh]
\begin{center}
\includegraphics[scale=0.35]{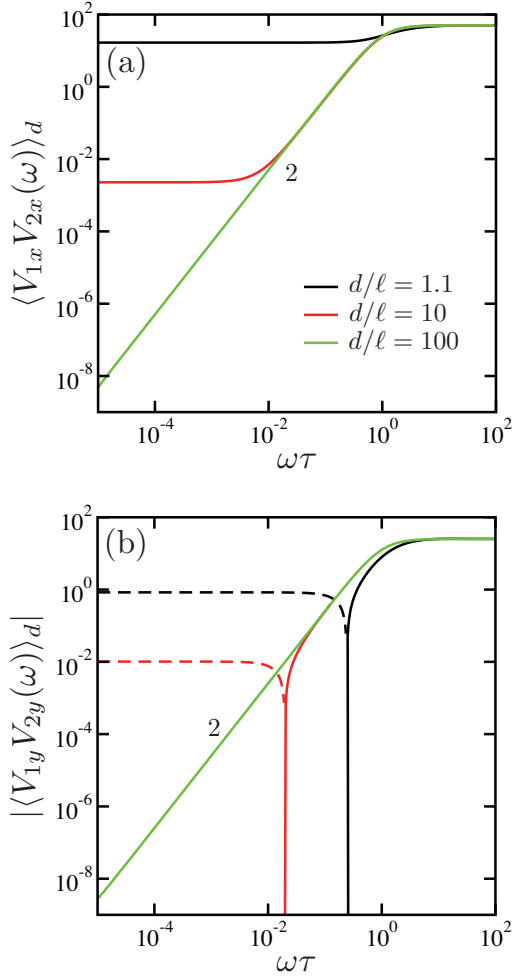}
\end{center}
\caption{
The active component of the scaled power spectral density (PSD) 
(a) $\langle V_{1x}V_{2x}(\omega)\rangle_d$ and 
(b) $\vert \langle V_{1y}V_{2y}(\omega)\rangle_d \vert$ [see Eq.~(\ref{active2psd})] 
as a function of $\omega \tau$ for $d/\ell=1.1$, $10$, $100$.
Here $d$ is the distance between the two point particles immersed in viscoelastic media.
Both PSDs are scaled by $c_0 \langle m^2(\omega)\rangle/(960\pi^2\eta^2 d)$ 
in order to make them dimensionless. 
Since $\langle V_{1y}V_{2y}(\omega)\rangle_d$ takes negative values for smaller $\omega \tau$
(shown by the dashed lines), we have plotted in (b) its absolute value.
The numbers indicate the slope representing the exponent of the power-law behaviors.
}
\label{2pointPSD}
\end{figure}

\begin{figure}[tbh]
\begin{center}
\includegraphics[scale=0.35]{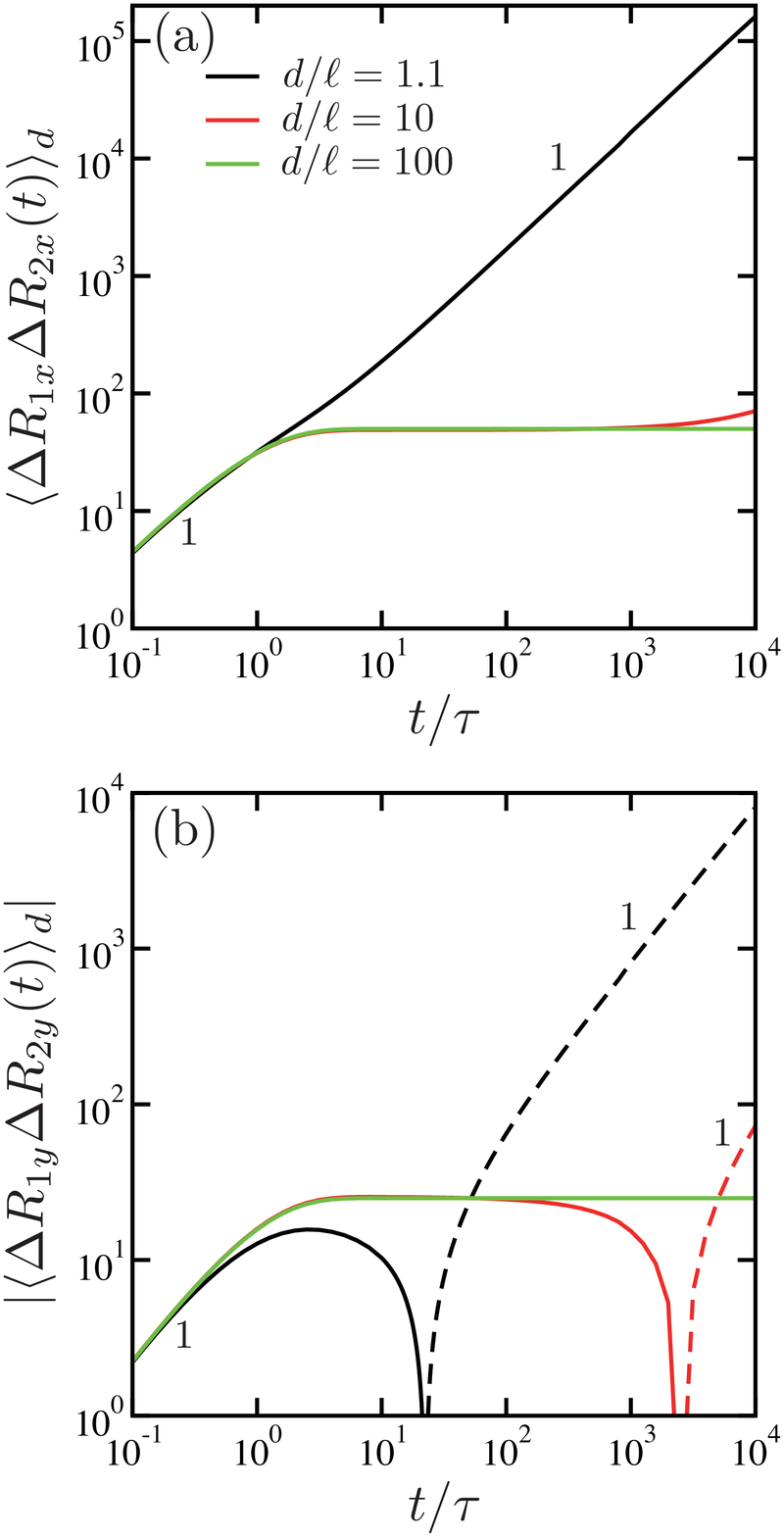}
\end{center}
\caption{
The active component of the two-point displacement cross-correlation functions (CCFs) 
(a) $\langle \Delta R_{1x} \Delta R_{2x}(t)\rangle_d$ and 
(b) $\vert \langle \Delta R_{1y} \Delta R_{2y}(t)\rangle_d \vert$ 
as a function of $t/\tau$ for $d/\ell=1.1$, $10$, $100$.
Here $d$ is the distance between the two point particles immersed in viscoelastic media, 
and the time correlation of the force dipole is $\delta$-correlated [see Eq.~(\ref{deltacorr})].
Both CCFs are scaled by $c_0 S\tau/(960 \pi^2\eta^2d)$
in order to make them dimensionless. 
Since $\langle \Delta R_{1y} \Delta R_{2y}(t)\rangle_d$ takes negative values for larger 
$t/ \tau$ (shown by the dashed lines), we have plotted in (b) its absolute value.
The numbers indicate the slope representing the exponent of the power-law behaviors.
}
\label{2pointMSDnoneq0}
\end{figure}

\begin{figure}[tbh]
\begin{center}
\includegraphics[scale=0.35]{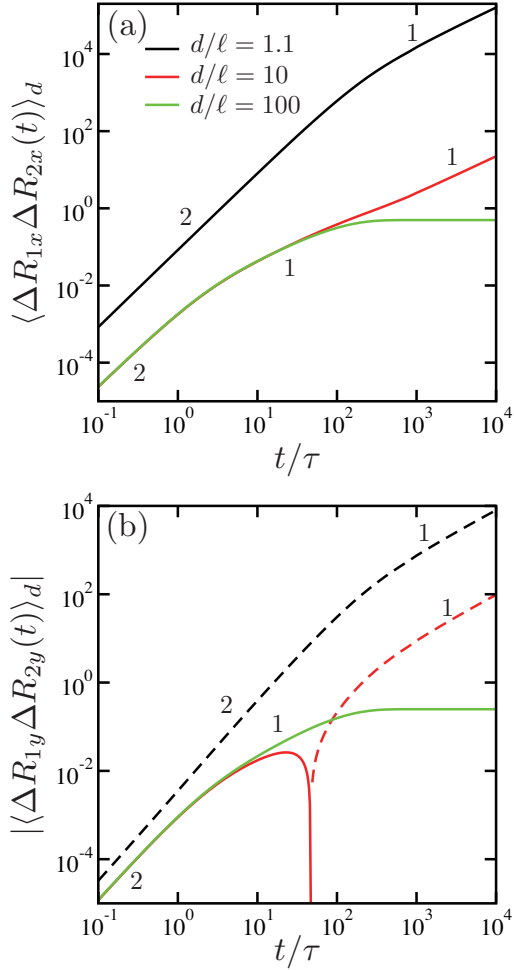}
\end{center}
\caption{
The active component of the two-point displacement cross-correlation functions (CCFs) 
(a) $\langle \Delta R_{1x} \Delta R_{2x}(t)\rangle_d$ and 
(b) $\vert \langle \Delta R_{1y} \Delta R_{2y}(t)\rangle_d \vert$ 
as a function of $t/\tau$ for $d/\ell=1.1$, $10$, $100$.
Here $d$ is the distance between the two point particles immersed in viscoelastic media.
The time correlation of the force dipole decays exponentially with a characteristic relaxation 
time $\tau_{\rm d}$ [see Eq.~(\ref{expcorr})], and we set here $\tau_{\rm d}/\tau=100$.
Both CCFs are scaled by $c_0 S\tau/(960 \pi^2\eta^2d)$
in order to make them dimensionless. 
Since $\langle \Delta R_{1y} \Delta R_{2y}(t)\rangle_d$ takes negative values for larger 
$t/ \tau$ (shown by the dashed lines), we have plotted in (b) its absolute value.
The numbers indicate the slope representing the exponent of the power-law behaviors.
}
\label{2pointMSDnoneq100}
\end{figure}

As before, we can generally set $\mathbf{R}_{2}-\mathbf{R}_{1}=d \hat{\mathbf{e}}_x$
without loss of generality.
We also assume that the active force dipoles are uniformly distributed in space with a constant 
concentration, $c_0$. 
Then one can further rewrite as  
\begin{align}
\langle V_{1\alpha}V_{2\alpha'}(\omega)\rangle_d= 
\frac{1}{8^2 \cdot 15\pi^2}\frac{c_0}{\eta^2\ell}\langle m^2(\omega)\rangle
\mathcal{I}_{\alpha\alpha'}(d, \omega),
\label{active2psd}
\end{align}
where
\begin{align}
& \mathcal{I}_{\alpha\alpha'}(d,\omega)=15\Omega_{\beta\beta'\gamma\gamma'}
\nonumber \\
& \times \int {\rm d}^3 \bar{r}\, \frac{\partial g_{\alpha\beta}[\mathbf{r},\omega]}{\partial \bar{r}_\gamma}
\left(\frac{\partial g_{\alpha'\beta'}[\mathbf{r}',-\omega]}
{\partial \bar{r}_{\gamma'}}\right)_{\mathbf{r}'=\mathbf{r}-\mathbf{d}},
\end{align}
with $g_{\alpha\beta}=8\pi\eta\ell G_{\alpha\beta}$ as defined before.

In Fig.~\ref{2pointPSD},  we numerically plot the scaled active PSDs 
$\langle V_{1x}V_{2x}(\omega)\rangle_d$ and 
$ \langle V_{1y}V_{2y}(\omega)\rangle_d$ as a function of $\omega \tau$ for 
different distances $d/\ell=1.1$, $10$ and $100$, as before.
(The reason that we chose here $d/\ell=1.1$ is that there was a numerical stability issue 
exactly at $d/\ell=1$.)
The PSD increases as $\sim \omega^2$ for the intermediate frequency range. 
Within the lowest order term in Eq.~(\ref{eq:Glong}), the asymptotic expressions of the 
active PSDs can be obtained as 
\begin{align}
\mathcal{I}_{xx}(d, \omega)\approx\frac{16\pi \ell}{d}\frac{(\omega\tau)^2}{1+ (\omega \tau)^2}, 
\end{align}
\begin{align}
\mathcal{I}_{yy}(d, \omega)=\mathcal{I}_{zz}(d, \omega)\approx
\frac{8\pi \ell}{d}\frac{(\omega\tau)^2}{1+ (\omega \tau)^2}. 
\end{align}

\subsection{Displacement cross-correlation functions}

Performing the inverse Fourier transform of the active two-point PSDs as before 
[see Eq.~(\ref{msdtimefourier})], we obtain the corresponding longitudinal and 
transverse displacement CCFs $\langle \Delta R_{1x} \Delta R_{2x}(t)\rangle_d$ 
and $\langle \Delta R_{1y} \Delta R_{2y}(t)\rangle_d$ for the distances $d/\ell=1.1$,
$10$ and $100$. 
In Fig.~\ref{2pointMSDnoneq0}(a) and (b),  we plot these quantities when the time correlation of 
a force dipole is $\delta$-correlated as assumed in Eq.~(\ref{deltacorr}). 
Fig.~\ref{2pointMSDnoneq0} should be compared with Fig.~\ref{1pointMSD}(a) where we have 
shown the MSD for the active one-point case.
Both longitudinal and transverse displacement CCFs are proportional to $t$ for 
short time scales $t/\tau < 1$ and also for longer time scales. 
For the intermediate time range, however, these CCFs are strongly suppressed and become
constant due to the elastic component of the two-fluid model.

In Fig.~\ref{2pointMSDnoneq100}(a) and (b), on the other hand, we consider the case when 
the time correlation of a force dipole is characterized by a relaxation time 
$\tau_{\rm d}/\tau=100$ [see Eq.~(\ref{expcorr})]. 
These figures should be compared with Fig.~\ref{1pointMSD}(b) because the overall behavior 
is similar. 
For $\delta/\ell=1.1$, the active displacement CCFs are proportional to $t^2$ when 
$t < \tau_{\rm d}$, showing a strong super-diffusive behavior.
For $\delta/\ell=100$, however, this super-diffusive behavior is observed only within the 
time region smaller than the viscoelastic time scale, $t/\tau <1$, and the CCFs increase 
as $\sim t$ for $t/\tau>1$. 
For much longer time scales, the active CCFs are further suppressed because of the elasticity.
In the long time limit, the active CCFs are both proportional to $t$.

\section{Summary and discussion}
\label{sec:discussion}

In this paper, we have discussed anomalous diffusion induced by active force dipoles in 
viscoelastic media that is described by the standard two-fluid model for gels. 
We first reviewed the two-fluid model and showed its partial Green's function both in the Fourier 
and the real spaces. 
With the use of the coupling mobilities and the FDT in thermal equilibrium, we have calculated 
the PSD of the velocity CCFs and the displacement CCFs between the two point particles both 
for the longitudinal and the transverse directions. 
The obtained results are useful to interpret the data obtained by two-point microrheology 
experiments. 
The passive (thermal) two-point CCF increases linearly with time at shorter and longer 
time scales, while it is suppressed and remains almost constant at intermediate time scales
(see Fig.~\ref{2pmsd}).

Moreover, we have calculated active (non-thermal) one-point and two-point 
correlation functions due to active force dipoles. 
We have used the relation between the velocity and the dipole strength, as given by 
Eq.~(\ref{valphart}), and the formulation in Ref.~\cite{Mikhailov15} in order to further 
calculate the active PSD of the velocity CCFs. 
For the one-point case, one needs to introduce a cutoff length scale, $\delta$, in evaluating 
the integrals, whereas a finite distance, $d$, between the two point particles plays the role 
of the cutoff length in the two-point case. 
As for the statistical property of force dipoles, we considered the case when their magnitude 
is uncorrelated in time [see Eq.~(\ref{deltacorr})] and the case when it decays exponentially 
with a characteristic time $\tau_{\rm d}$ [see Eq.~(\ref{expcorr})].

For the active case, the important results can be summarized as follows.
As shown in Fig.~\ref{1pointMSD}(b) (one-point case) or Fig.~\ref{2pointMSDnoneq100} 
(two-point case), we have found that the active MSD or the displacement CCFs exhibits 
various crossovers from super-diffusive to sub-diffusive behaviors depending on the 
characteristic time scales ($\tau=\eta/\mu$ and $\tau_{\rm d}$) and the particle 
separation $d$ (or the cutoff length $\delta$ for the one-point case).
We emphasize that the active displacement CCF is proportional to $t^2$ for time scales 
shorter than the viscoelastic time scale, $t < \tau$, and it is proportional to $t$ for the 
intermediate time scales,  $\tau < t < \tau_{\rm d}$.
Within the present model, the passive contribution only describes sub-diffusion, whereas 
the active contribution is responsible for both sub-diffusion and super-diffusion.
Our results are useful in understanding active properties of the cytoplasm using force 
spectrum microscopy combined with the microrheology experiment~\cite{Guo14}, 
as further discussed below.

In Ref.~\cite{Guo14}, Guo \textit{et al}.\ measured the MSD of microinjected tracer 
particles in mellanoma cells.
They showed that the MSD was nearly constant at shorter time scales ($t < 0.1$~s), while 
it exhibited a slightly super-diffusive behavior at longer time scales ($t > 0.1$~s), i.e., 
$\langle (\Delta R)^2 \rangle \sim t^{\beta}$ with $\beta \approx 1.2$. 
However, when they inhibited motor and polymerization activity by depleting cells of ATP,
the MSD was almost constant in time, i.e., $\beta \approx 0$.
Such an ATP-dependent Brownian motion was also observed in prokaryotic cells and 
yeast~\cite{Weber12,Parry14}.
In addition to the MSD measurement, Guo \textit{et al}.\ performed active microrheology 
experiment~\cite{GSOMS,Schnurr97}, and found that the frequency-dependent elastic modulus
follows a power-law form, i.e., $\vert G(\omega) \vert \sim \omega^{\alpha}$
with $\alpha \approx 0.15$~\cite{Guo14}.  

For simplicity, one may assume that PSD of the active force also obeys a power-law behavior, 
i.e., $\langle m^2 (\omega)\rangle \sim \omega^{-\gamma}$ with a different exponent 
$\gamma$.
Among these three exponents, the following scaling relation should hold~\cite{Guo14,Lau03}:
\begin{equation}
\beta=2 \alpha+\gamma-1.
\label{scaling}
\end{equation}
In thermal equilibrium, $\gamma= - \alpha + 1$ holds according to the FDT and hence $\beta=\alpha$.
In this case, the anomalous diffusion purely reflects the viscoelasticity of the surrounding media.
The exponentially correlated force dipoles in Eq.~(\ref{expcorr}) leads to the active PSD in 
Eq.~(\ref{activePSD}), and hence $\gamma=2$ for $\omega \tau_{\rm d} \gg 1$. 
Experimentally, the value $\gamma \approx 2$ was observed by Lau 
\textit{et al}.~\cite{Lau03} and later reconfirmed by Guo \textit{et al}.~\cite{Guo14}. 
When $\alpha \approx 0.15$ and $\gamma \approx 2$, Eq.~(\ref{scaling}) gives 
$\beta \approx 1.3$ which is almost consistent with the MSD measurement mentioned above
($\beta \approx 1.2$). 
In the older experiment~\cite{Lau03}, on the other hand, the measured exponents were 
$\alpha \approx 0.25$, $\gamma \approx 2$ and hence $\beta \approx 1.5$.
In both of these experiments, they claimed that active forces dominate the low-frequency regime,  
whereas thermal forces dominate the high-frequency regime~\cite{Guo14,Lau03}.

It should be reminded, however, that different values of $\gamma$ were 
reported by different groups~\cite{Wilhelm08,Gallet09,Robert10}.
For example, a combination of active and passive microrheology measurements using 
PC3 tumor cells resulted in $\alpha \approx 0.4$, $\beta \approx 1.3$ and 
$\gamma \approx 1.5$, satisfying also the scaling relation Eq.~(\ref{scaling}).  
They argued that such a difference can arise because active and passive measurements 
were done in Ref.~\cite{Lau03} with different probes and at very different locations in 
the cell. 
In Refs.~\cite{Wilhelm08,Gallet09,Robert10}, they performed dual 
passive-active measurements with a unique probe.
Given these situations, we consider that the power-law behavior of the force fluctuations
and its exponent require further experimental and theoretical investigations.

We also point out that Eq.~(\ref{scaling}) cannot be always true because the exponents 
can take values only $0 \le \beta \le 2$ and $0 \le \alpha \le 1$. 
Hence, if $\gamma = 2$ holds, one should always observe a super-diffusive behavior 
because $\beta = 2 \alpha + 1 \ge 1$. 
However, sub-diffusive behaviors ($\beta < 1$) in cells have been observed in many 
cases~\cite{Norrelykke10,Golding06, Weber10}.
Moreover, the above relation also restricts the value of $\alpha$ to $0< \alpha \le 0.5$ 
because $\beta <2$, which is not always the case~\cite{Hoffman06,Hoffman09}.

In our work, we have assumed that the time correlation of a force dipole is an exponentially 
decaying function with a characteristic time $\tau_{\rm d}$, as given in Eq.~(\ref{expcorr}).
Hence its Fourier transform has a Lorentzian form, and decays as $\omega^{-2}$ for 
$\omega \tau_{\rm d} \gg 1$. 
A similar Lorentzian form of force fluctuations was discussed by Levine and 
MacKintosh~\cite{Mac08,Levine09}.
While some of the experiments which reported the exponent 
$\gamma=2$~\cite{Guo14,Lau03} justify our assumption, different values of $\gamma$ found 
in the other experiments~\cite{Wilhelm08,Gallet09,Robert10} indicate 
that the dipole correlation cannot be a simple exponentially decaying function.
Hence a more detailed investigation for the statistical property of a fluctuating  force dipole 
is required.
Currently, we are analyzing the stochastic properties of a simple model of a catalytically active 
bidomain protein~\cite{Mikhailov15}. 
In this model, the two protein domains are represented by beads connected by an elastic spring, 
and the two internal states, namely, free protein and ligand-protein complex, are assumed.

Although our theory is general and can be applied not only for cells but also for other 
macroscopic systems, it is useful to give some typical parameter values corresponding to a cell.
Since the characteristic length scale $\ell=(\eta/\Gamma)^{1/2}$ roughly corresponds to 
the mesh size of a polymer gel, it is roughly given by $\ell \sim 10^{-7}$~m for a typical cell.
Hence the distance between the two point particles such as $d/\ell=100$ means  
$d \sim 10^{-5}$~m.
According to Ref.~\cite{Guo14}, we also have $\eta\sim 10^{-3}$~Pa$\cdot$s and $\mu\sim 1$~Pa
so that the viscoelastic time scale can be estimated as $\tau=\eta/\mu \sim 10^{-3}$~s.
This means that the dipole time scale $\tau_{\rm d}/\tau=100$ used in 
Figs.~\ref{1pointMSD}(b) and \ref{2pointMSDnoneq100} corresponds to 
$\tau_{\rm d}\sim 10^{-1}$~s.
Although this time scale is somewhat larger than the cycle time of a single protein 
machine~\cite{Mikhailov15}, it still gives a good estimate to characterize the collective 
dynamics of a protein complex.

Recently, Fodor \textit{et al.}~\cite{Fodor15} have made an attempt to theoretically 
reproduce the MSD data measured in the cytoplasm of living A7~\cite{Guo14}.
They used one-dimensional Langevin equation in the presence of a random active force
to calculate both the thermal and non-thermal MSD. 
Their theory has a similarity to the present work because they also introduce two 
time scales which are analogous to  $\tau=\eta/\mu$ and $\tau_{\rm d}$ in our theory.
An important new aspect in the present paper is that the internal structure of the 
viscoelastic medium is properly taken into account.  
Both thermal and non-thermal MSDs exhibit complicated time sequences depending 
on the length-scale of the observation relative to the mesh size $\ell$.
In Ref.~\cite{Fodor15}, the size of tracer particles was assumed to be always 
larger than the mesh size of the cytoskeletal network.

In our separate work, we have considered the two-fluid model where active macromolecules, 
described as force dipoles, cyclically operate both in the elastic and the fluid 
components~\cite{Yasuda17}. 
Through coarse-graining, we have derived effective equations of motions for tracer particles 
displaying local deformations and local fluid flows. 
The equation for deformation tracers coincides with the phenomenological model by 
Fodor \textit{et al.}~\cite{Fodor15} (see also the related publication~\cite{Fodor16}).
Our analysis reveals that localization and diffusion phenomena are generally involved. 
The motion of tracers immobilized within the elastic subsystem is localized in the long-time limit, 
but it can show a diffusion-like behavior at the intermediate time scales shorter than the cooperative 
correlation times of molecular motor aggregates operating in the active gels~\cite{Yasuda17}.

Recently, Bruinsma \textit{et al.}~\cite{Bruinsma14} investigated a large scale correlated 
motion of chromatin inside the nuclei of living cells by using another ``two-fluid model" 
for polymer solutions~\cite{Doi92} (but not for gels).
They derived the response functions that connect the chromatin density and 
velocity correlation function to the correlation functions of the active sources that 
are either scalar or vector quantities.
One of the differences in their theory is that the form of the complex viscoelastic moduli 
needs to be specified in order to compare with experiments, whereas the viscoelasticity
naturally arises from the present two-fluid model.  
It is interesting to note that their active PSD also contains Lorentzian type frequency 
dependence as we have obtained such as in Eq.~(\ref{activePSD}). 
It would be interesting to calculate the active MSD based on this different two-fluid model.

Finally, we mention that anomalous diffusion observed in colloidal gels has been also 
explained in terms of force dipoles due to structural inhomogeneities~\cite{Cipelletti,Duri06}.
Assuming that such inhomogeneities are randomly distributed, it was shown that the relaxation 
time of the dynamic structure factor is inversely proportional to the wavenumber.
In Ref.~\cite{Gao}, the MSD exhibits diffusive motion at short times and super-diffusive motion 
at long times.

\begin{acknowledgments}

We thank A.\ S.\ Mikhailov and T.\ Kato for stimulating discussions.
S.K. acknowledges support from the Grant-in-Aid for Scientific Research on
Innovative Areas ``\textit{Fluctuation and Structure}" (Grant No.\ 25103010) from the Ministry
of Education, Culture, Sports, Science, and Technology of Japan, and the 
Grant-in-Aid for Scientific Research (C) (Grant No.\ 15K05250)
from the Japan Society for the Promotion of Science (JSPS).
\end{acknowledgments}

\appendix
\section{Partial Green's function}
\label{appgreenfunction}

In this appendix, we show the derivation of Eqs.~(\ref{eq:oseen_q}) and 
(\ref{fullgreen})~\cite{Levine00,Levine01,Sonn14,Sonn14-2}.
By using the Fourier transform in space and the Laplace transform in time, Eqs.~(\ref{model1}), 
(\ref{model2}) and (\ref{incompress}) can be represented in the steady state as 
\begin{align}
0 =& -\mu q^2 \mathbf{u}[\mathbf{q},s]-(\mu+\lambda)\mathbf{q}
(\mathbf{q}\cdot \mathbf{u}[\mathbf{q},s]) \nonumber \\
& -\Gamma \left(s\mathbf{u}[\mathbf{q},s]-\mathbf{v}[\mathbf{q},s] \right),
\label{model1_app}
\end{align}
\begin{align}
0=&-\eta q^2 \mathbf{v}[\mathbf{q},s] -i\mathbf{q} p[\mathbf{q},s] 
\nonumber \\
&-\Gamma \left(\mathbf{v}[\mathbf{q},s]- s\mathbf{u}[\mathbf{q},s]\right)+\mathbf{f}_v[\mathbf{q},s],
\label{model2_app}
\end{align}
\begin{align}
\mathbf{q}\cdot\mathbf{v}[\mathbf{q},s]=0.
\label{incompress_app}
\end{align}
Taking the inner products of both Eqs.~(\ref{model1_app}) and (\ref{model2_app}) with 
$\mathbf{q}$, and using Eq.~(\ref{incompress_app}), we obtain 
\begin{align}
p[\mathbf{q},s]=-\frac{i\mathbf{q}\cdot\mathbf{f}_v[\mathbf{q},s]}{q^2}.
\label{p_app}
\end{align}

From Eq.~(\ref{model1_app}), we can solve for $\mathbf{u}$ as 
\begin{align}
u_\alpha[\mathbf{q},s]&=\left(\frac{\Gamma}{\mu q^2+s\Gamma}\delta_{\alpha\beta}  \right. \nonumber \\
& \left. 
-\frac{\Gamma(\mu+\lambda)q^2}{(\mu q^2+s\Gamma)(2\mu q^2+\lambda q^2+s\Gamma)}
\hat q_\alpha\hat q_\beta\right)v_\beta[\mathbf{q},s].
\label{u_app}
\end{align}
Substituting Eqs.~(\ref{p_app}) and (\ref{u_app}) into Eq.~(\ref{model2_app}), we obtain
the following equation 
\begin{align}
\biggl(& \frac{q^2[\eta\mu q^2+(s\eta+\mu)\Gamma]}{\mu q^2+s\Gamma}\delta_{\alpha\beta}
\nonumber \\
& -\frac{s\Gamma^2(\mu+\lambda)q^2}
{(\mu q^2+s\Gamma)(2\mu q^2+\lambda q^2+s\Gamma)}\hat q_\alpha\hat q_\beta \biggr)
v_\beta[\mathbf{q},s] \nonumber \\
=&(\delta_{\alpha\beta}-\hat q_\alpha\hat q_\beta)f_{v,\beta}[\mathbf{q},s].
\end{align}
Then we can solve for $\mathbf{v}$ as 
\begin{align}
v_\alpha[\mathbf{q},s]=\frac{\mu q^2+s\Gamma}{q^2[\eta\mu q^2+(s\eta+\mu)\Gamma]}
(\delta_{\alpha\beta}-\hat q_\alpha\hat q_\beta)f_{v,\beta}[\mathbf{q},s].
\end{align}
In terms of $\eta_{\rm b}$ and $\xi$ defined in Eq.~(\ref{xiomega}), we finally obtain 
Eq.~(\ref{eq:oseen_q}).

Next, we derive the real space representation of the partial Green's function~\cite{Doi}.
We first assume that it has the form of
\begin{equation}
G_{\alpha\beta}[\mathbf{r},s]=C_1\delta_{\alpha \beta}+C_2\hat{r}_{\alpha} \hat{r}_{\beta},
\end{equation}
so that 
\begin{equation}
G_{\alpha \alpha}[\mathbf{q},s]=3C_1+C_2,
\label{eq:a1}
\end{equation}
\begin{equation}
G_{\alpha \beta}[\mathbf{q},s]\hat{r}_{\alpha} \hat{r}_{\beta}=C_1+C_2.
\label{eq:b1}
\end{equation}
Hence we have 
\begin{align}
3C_1+C_2 & = 2\int \frac{d^3q}{(2\pi)^3}\,\frac{1+(\eta_{\rm b}/\eta)\xi^2q^2}
{\eta_{\rm b} q^2(1+\xi^2q^2)}
e^{i\mathbf{q} \cdot \mathbf{r}},
\nonumber \\
&=\frac{1}{2\pi\eta r}\left[1+\frac{1-\eta_{\rm b}/\eta}{\eta_{\rm b}/\eta}
(1-e^{-r/\xi})\right],
\end{align}
\begin{align}
C_1+C_2&=\int \frac{d^3q}{(2\pi)^3}\,\frac{1+(\eta_{\rm b}/\eta)\xi^2q^2}
{\eta_{\rm b} q^2(1+\xi^2q^2)}\left[1-(\hat{\mathbf{q}}\cdot \hat{\mathbf{r}})^2\right]
e^{i\mathbf{q} \cdot \mathbf{r}} 
\nonumber \\
& =\frac{1}{4\pi\eta r}\biggl[1+\frac{1-\eta_{\rm b}/\eta}{\eta_{\rm b}/\eta}
\Bigl(1-2(\xi/r)^2
\nonumber \\
&+2 e^{-r/\xi}\bigl[(\xi/r)+(\xi/r)^2\bigr] \Bigr)\biggr].
\end{align}
Solving for $C_1$ and $C_2$, we finally arrive at Eq.~(\ref{fullgreen}) with Eqs.~(\ref{G1(x)})  
and (\ref{G2(x)}).

\section{Derivation of Eq.~(\ref{eq:spectol_high})}
\label{appderivepsd}

In this appendix, we show the derivation of Eq.~(\ref{eq:spectol_high}).
Here we use the dimensionless form of the Green's function  
$g_{\alpha\beta}=8\pi\eta\ell G_{\alpha\beta}$, and consider its 
asymptotic expression [see Eq.~(\ref{eq:Glong})]
\begin{align}
g_{\alpha\beta}[\mathbf{r},\omega]&=\frac{i\omega\tau\ell}{r(1+i\omega\tau)}
(\delta_{\alpha \beta}+\hat{r}_{\alpha} \hat{r}_{\beta})  
\nonumber \\
& -\frac{2\ell^3}{r^3(1+i\omega\tau)^2}
(\delta_{\alpha \beta}-3\hat{r}_{\alpha} \hat{r}_{\beta})
\nonumber \\
&\equiv A_{\alpha\beta}(\mathbf{r},\omega)-B_{\alpha\beta}(\mathbf{r},\omega), 
\end{align}
where we have defined the functions $A_{\alpha\beta}$ and $B_{\alpha\beta}$ in the last equation.
The spatial derivatives of these functions with respect to $\bar{r}=r/\ell$ are 
\begin{align}
& \frac{\partial}{\partial \bar r_\gamma}A_{\alpha\beta}(\mathbf{r},\omega)=
\frac{i\omega\tau}{1+i\omega\tau}
\nonumber \\
& \times \left(\frac{\bar r_\alpha\delta_{\beta\gamma}+\bar r_\beta\delta_{\alpha\gamma}-
\bar r_\gamma\delta_{\alpha\beta}}{\bar r^3}-3\frac{\bar r_\alpha \bar r_\beta \bar r_\gamma}{\bar r^5}\right),
\end{align}
\begin{align}
&\frac{\partial}{\partial \bar r_\gamma}B_{\alpha\beta}(\mathbf{r},\omega)=
\frac{2}{(1+i\omega\tau)^2}
\nonumber \\
& \times \left(-3\frac{\bar r_\alpha\delta_{\beta\gamma}+\bar r_\beta\delta_{\alpha\gamma}+\bar r_\gamma\delta_{\alpha\beta}}{\bar r^5}+15\frac{\bar r_\alpha \bar r_\beta \bar r_\gamma}{\bar r^7}\right).
\end{align}

Using these results, we can further calculate the following quantities
\begin{align}
& 15\Omega_{\beta\beta'\gamma\gamma'}
\frac{\partial}{\partial \bar r_\gamma}A_{\alpha\beta}(\mathbf{r},\omega)
\frac{\partial}{\partial \bar r_{\gamma'}}A_{\alpha\beta'}(\mathbf{r},-\omega)
\nonumber \\
&=12\frac{(\omega\tau)^2}{1+(\omega\tau)^2}\bar r^{-4},
\end{align}
\begin{align}
&15\Omega_{\beta\beta'\gamma\gamma'}
\frac{\partial}{\partial \bar r_\gamma}A_{\alpha\beta}(\mathbf{r},\omega)
\frac{\partial}{\partial \bar r_{\gamma'}}B_{\alpha\beta'}(\mathbf{r},-\omega)
\nonumber \\
&=-72\frac{i\omega\tau-(\omega\tau)^2}{[1+(\omega\tau)^2]^2}\bar r^{-6},
\end{align}
\begin{align}
&15\Omega_{\beta\beta'\gamma\gamma'}
\frac{\partial}{\partial \bar r_\gamma}B_{\alpha\beta}(\mathbf{r},\omega)
\frac{\partial}{\partial \bar r_{\gamma'}}A_{\alpha\beta'}(\mathbf{r},-\omega)
\nonumber \\
&= -72\frac{-i\omega\tau-(\omega\tau)^2}{[1+(\omega\tau)^2]^2}\bar r^{-6},
\end{align}
\begin{align}
&15\Omega_{\beta\beta'\gamma\gamma'}
\frac{\partial}{\partial \bar r_\gamma}B_{\alpha\beta}(\mathbf{r},\omega)
\frac{\partial}{\partial \bar r_{\gamma'}}B_{\alpha\beta'}(\mathbf{r},-\omega)
\nonumber \\
& =\frac{720}{[1+(\omega\tau)^2]^2}\bar r^{-8}.
\end{align}
Hence the dimensionless PSD in Eq.~(\ref{Iomega}) can be obtained as  
\begin{align}
\mathcal{I}(\omega)&= 15\Omega_{\beta\beta'\gamma\gamma'}\int {\rm d}^3\bar r\, \frac{\partial g_{\alpha\beta}[\mathbf{r},\omega]}{\partial \bar r_\gamma}
\frac{\partial g_{\alpha\beta'}[\mathbf{r},-\omega]}{\partial \bar r_{\gamma'}} \nonumber \\
&= 4\pi\int_{\delta/\ell}^\infty {\rm d}\bar r\,\bar r^2\left(12\frac{(\omega\tau)^2}{1+(\omega\tau)^2}\bar r^{-4}
\right.
\nonumber \\
& \left.  -144\frac{(\omega\tau)^2}{[1+(\omega\tau)^2]^2}\bar r^{-6}+\frac{720}{[1+(\omega\tau)^2]^2}\bar r^{-8}\right) 
\nonumber \\
& = 48\pi\left[\frac{(\omega\tau)^2}{1+(\omega\tau)^2}\left(\frac{\ell}{\delta}\right)-4\frac{(\omega\tau)^2}{[1+(\omega\tau)^2]^2}\left(\frac{\ell}{\delta}\right)^3
\right.
\nonumber \\
& \left. +\frac{12}{[1+(\omega\tau)^2]^2}\left(\frac{\ell}{\delta}\right)^5\right].
\end{align}
Hence we finally obtain Eq.~(\ref{eq:spectol_high}).

\section{Full expression of Eq.~(\ref{msdlowestorder})}
\label{appmsdhigherorder}

The full expression of Eq.~(\ref{msdlowestorder}) including higher order terms 
in $\ell/\delta$ is given as follows
\begin{widetext}
\begin{align}
\langle (\Delta R_x)^2(t) \rangle 
&\approx  \frac{1}{60\pi}\frac{c_0S\tau}{\eta^2\ell}
\Biggl[
\biggl(\frac{1}{1+\tau_{\rm d}/\tau}
+\frac{(\tau_{\rm d}/\tau) e^{-t/\tau_{\rm d}}}{1-(\tau_{\rm d}/\tau)^2}-\frac{e^{-t/\tau}}{1-(\tau_{\rm d}/\tau)^2}\biggr) \left( \frac{\ell}{\delta}\right) \nonumber \\
&+\Biggl(\frac{\left[-2 (t/\tau) (\tau_{\rm d}/\tau)^2+2 t/\tau-6 (\tau_{\rm d}/\tau)^2+2\right]e^{-t/\tau} }{\left[1-(\tau_{\rm d}/\tau)^2\right]^2}+\frac{4 (\tau_{\rm d}/\tau)^3 e^{-t/\tau_{\rm d}}}{ \left[1-(\tau_{\rm d}/\tau)^2\right]^2}-\frac{2 (2 \tau_{\rm d}/\tau+1)}{[1+\tau_{\rm d}/\tau]^2}\Biggr)\left( \frac{\ell}{\delta}\right)^3 \nonumber \\
&+\Biggl(\frac{6 \left[-(t/\tau) (\tau_{\rm d}/\tau)^2+t/\tau-5 (\tau_{\rm d}/\tau)^2+3\right]e^{-t/\tau}}{\left[1-(\tau_{\rm d}/\tau)^2\right]^2}+\frac{12 (\tau_{\rm d}/\tau)^5 e^{-t/\tau_{\rm d}}}{ \left[1-(\tau_{\rm d}/\tau)^2\right]^2}\nonumber \\
&-\frac{6 [2 (\tau_{\rm d}/\tau)^3+4(\tau_{\rm d}/\tau)^2+6 \tau_{\rm d}/\tau+3]}{ [1+\tau_{\rm d}/\tau]^2} + \frac{12t}{\tau}\Biggl)\left( \frac{\ell}{\delta}\right)^5
\Biggr].
\end{align}
\end{widetext}


\end{document}